\documentclass[singlelespacing]{elsart}
\usepackage{epsfig}

\begin{document}

\begin{frontmatter}
\title{Semi-empirical calculation of quenching factors for ions in scintillators}

\author[] {V.I.~Tretyak\thanksref{1}}

\thanks[1]{Corresponding author.
           Address: Institute for Nuclear Research, Prospekt Nauki 47,
           MSP 03680 Kyiv, Ukraine;
           Telephone: +380 44 525 2210;
           Fax: +380 44 525 4463;
           E-mail address: tretyak@kinr.kiev.ua (Vladimir Tretyak).}

\address[KINR]{Institute for Nuclear Research, MSP 03680 Kyiv, Ukraine}
\address[SNU]{Department of Physics and Astronomy, Seoul National University,
              151-747 Seoul, Republic of Korea}

\begin{abstract}

Semi-empirical method of calculation of quenching factors for scintillators is described.
It is based on classical Birks formula with the total stopping powers for
electrons and ions which are calculated with the ESTAR and SRIM codes, respectively.
Method has only one fitting parameter (the Birks factor $kB$) which can have
different values for the same material in different conditions of measurements and data
treatment. A hypothesis is used that, once the $kB$ value is
obtained by fitting data for particles of one kind and in some
energy region (e.g. for a few MeV $\alpha$ particles from internal contamination of 
a detector), it can be applied to calculate quenching factors
for particles of another kind and for another energies (e.g. for low energy nuclear
recoils) if all data are measured in the same experimental conditions and are
treated in the same way.
Applicability of the method is demonstrated on many examples including materials
with different mechanisms of scintillation:
organic scintillators (solid C$_8$H$_8$, and liquid C$_{16}$H$_{18}$, C$_9$H$_{12}$);
crystal scintillators (pure CdWO$_4$, PbWO$_4$, ZnWO$_4$, CaWO$_4$, CeF$_3$,
and doped CaF$_2$(Eu), CsI(Tl), CsI(Na), NaI(Tl));
liquid noble gases (LXe).
Estimations of quenching factors for nuclear recoils are also given for some
scintillators where experimental data are absent (CdWO$_4$, PbWO$_4$, CeF$_3$, 
Bi$_4$Ge$_3$O$_{12}$, LiF, ZnSe).

\end{abstract}


\end{frontmatter}

\section{Introduction}

In accordance with our current understanding of astronomical observations,
usual matter constitutes only $\simeq4$\% of the Universe; the main components
are dark matter ($\simeq23$\%) and dark energy ($\simeq73$\%) \cite{Ber05}.
Various extensions of the Standard Model propose many candidates on the role
of dark matter (DM) particles \cite{Ste09} which are neutral and only weakly
interact with matter (Weakly Interacting Massive Particles, WIMPs).
One of the approaches to discover these particles is to detect scattering of
WIMPs on atomic nuclei in sensitive detectors placed deep underground
and measured in extra low background conditions \cite{Spo07}.
Taking into account likely mass range and velocities of WIMPs,
energies of nuclear recoils are expected below $\simeq100$ keV
with character interaction rates of $1-10^{-6}$ events kg$^{-1}$ d$^{-1}$.
Many searches of WIMPs with semiconductor, scintillator and bolometer
detectors to-date gave only negative results (see \cite{Spo07} and references
therein);
instead positive evidence for DM particles (WIMPs are a subclass; 
other candidates and other kinds of interactions are also available) 
in the galactic halo has been pointed out by DAMA experiments by exploiting 
the DM annual modulation signature with NaI(Tl) scintillators during more than 
10 years long measurements \cite{Ber08a}.

For a long time it is known that amount of light produced in scintillating material
by highly ionizing particles
is lower than that produced by electrons of the same energy \cite{Bir64}.
Thus, in a scintillator calibrated with electron and/or $\gamma$ sources
(which is an usual practice), signals from ions will be seen at lower energies
(sometimes up to $\simeq 40$ times) 
than their real values.
Evidently knowledge of these transformation coefficients
-- quenching factors -- is extremely important in
searches for WIMPs and in predictions where the WIMPs signal should be expected.
Many experimental efforts were devoted to measurements, sometimes very sophisticated,
of quenching factors at low energies in different detectors (see e.g.
recent works \cite{Bav08,Cal08,Cha08,Cor08,Kob08,Nik08,Sor09} and further references).

Quenching factors are also needed in measurements and interpretation of signals
from $\alpha$ particles in scintillators. As examples, we can mention here
recent experiments on searches (and first observations) of extremely rare
$\alpha$ decays ($T_{1/2}=10^{18}-10^{19}$ yr):
$^{180}$W in CdWO$_4$ \cite{Dan03} and CaWO$_4$ \cite{Zde05} crystal scintillators,
and $^{151}$Eu in CaF$_2$(Eu) \cite{Bel07}.

While few approaches in calculation of quenching factors are known
\cite{Bir51,Mur61,Lin63,Hit05}, satisfactory theory able to exactly predict
(and very often even to describe already measured)
quenching factors for all detectors and particles still is absent.
For example, in the Lindhard's approach \cite{Lin63}\footnote{Good description
is given in more accessible source \cite{Mei08}.} it is possible to calculate
quenching of ions with atomic number $Z$ in scintillator only
with the same $Z$ number; in addition, this theory predicts decrease of quenching
factors at low energies, very often in contradiction with experimental data.
Hitachi's model \cite{Hit05} gives better description and for wider data range,
however it is not easy to reproduce these calculations independently.

Below we describe rather simple method of calculation of quenching factors for
different ions (from protons to heavy recoils),
based on semi-empirical approach of Birks \cite{Bir51} and using available in Internet
software for calculation of stopping powers for electrons and ions (ESTAR \cite{ESTAR} and
SRIM \cite{SRIM} codes, respectively). It employs only one parameter ($kB$ Birks factor)
which could be found by fitting experimental data measured for particles
of one kind in some energy region (e.g. for $\alpha$ particles from external sources or
internal contamination of a detector by U/Th chains, $^{147}$Sm, $^{190}$Pt, etc.) 
but afterwards can be used to calculate
quenching factors for other particles and in other energy regions (e.g. for nuclear
recoils at low energies). Summary of the method is given in section 2.
Calculations with this method are demonstrated in section 3 for number of
scintillators:
organic scintillators (solid C$_8$H$_8$, and liquid C$_{16}$H$_{18}$, C$_9$H$_{12}$);
crystal scintillators (pure CdWO$_4$, PbWO$_4$, ZnWO$_4$, CaWO$_4$, CeF$_3$,
and doped CaF$_2$(Eu), CsI(Tl), CsI(Na), NaI(Tl));
liquid noble gases (LXe).
Estimations of quenching factors for nuclear recoils are also given for some
scintillators where experimental data are absent (CdWO$_4$, PbWO$_4$, CeF$_3$, 
Bi$_4$Ge$_3$O$_{12}$, LiF, ZnSe).
Section 4 gives conclusions.

\section{Outlines of the method}

In calculation of quenching factors, we follow Birks approach in
description of quenching of the light yield for highly ionizing
particles \cite{Bir64,Bir51}. Light yield of scintillating
material depends not only on energy of particle $E$ but also on
how big is its stopping power $dE/dr$ in the material. Fig. 1
gives example of stopping powers for electrons (calculated with
the ESTAR software \cite{ESTAR}), and for protons, alpha
particles, O, Ca and W ions (calculated with the SRIM code
\cite{SRIM}) in the CaWO$_4$ material.

\nopagebreak
\begin{figure}[htb]
\begin{center}
\mbox{\epsfig{figure=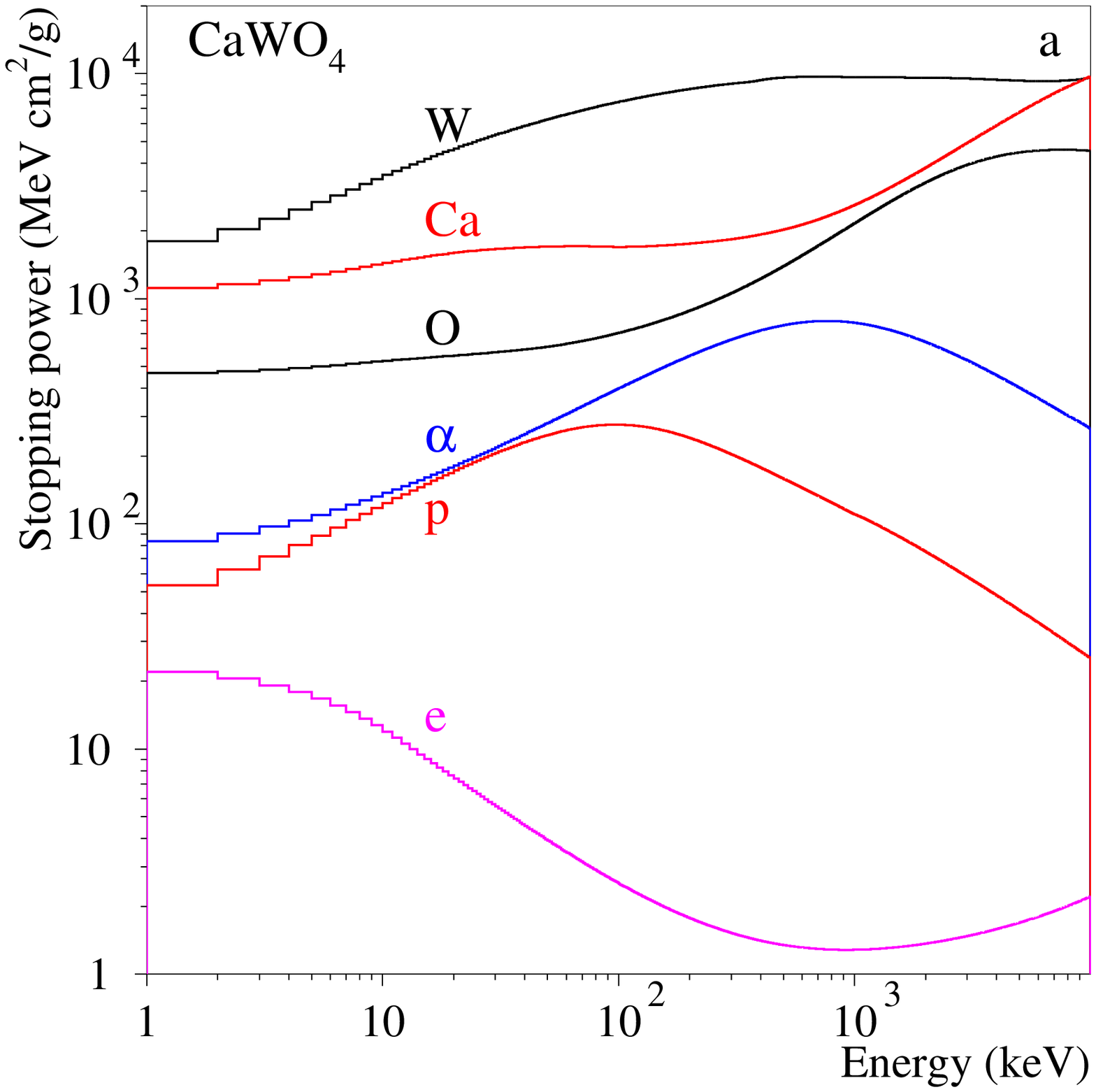,width=6.6cm}}~
\mbox{\epsfig{figure=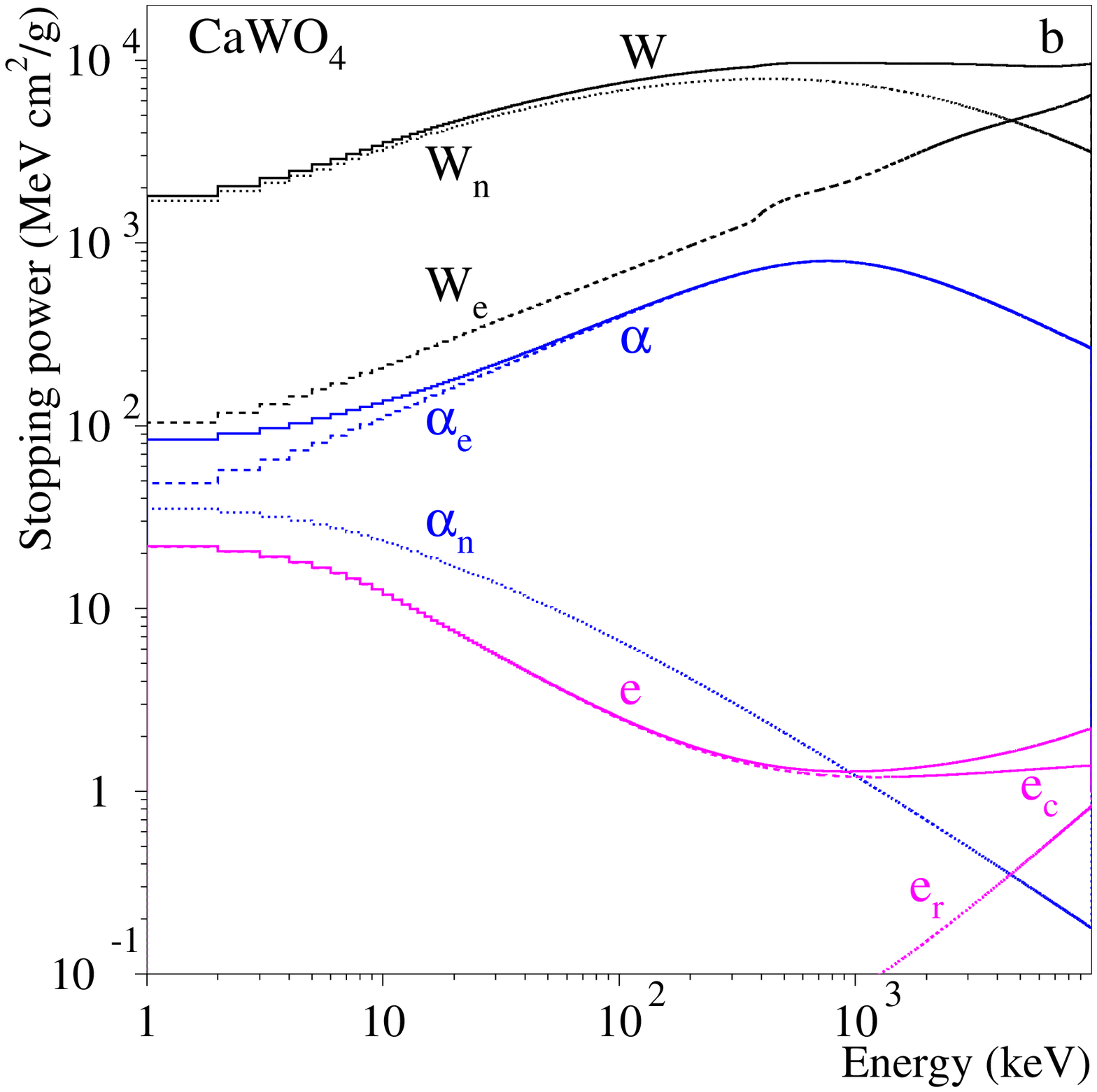,width=6.6cm}}
\caption{(Color online) Stopping powers for electrons, protons, $\alpha$ particles, O, Ca
and W ions in CaWO$_4$. In (a) only total $dE/dr$ are given; in (b) also
nuclear (dotted line) and electronic (dashed line) parts of $dE/dr$ are drawn
for $\alpha$ particles and W ions; for electrons, collision and radiation parts
are shown.}
\end{center}
\end{figure}

In case when created in a scintillator excitation centers are spaced at 
large distances and interactions between them can be neglected, 
what is realized for particles with low stopping power (fast electrons,
energies above $E \simeq 100$ keV),
scintillation yield $dL$ is proportional to released energy $dE$:
$dL = SdE$ (where $S$ is the absolute scintillation factor), 
or in differential form

\begin{equation}
\frac{dL}{dr} = S \frac{dE}{dr}.
\end{equation}

To account for suppression of the light yield for highly ionizing
particles (protons, $\alpha$ particles and nuclear recoils;
hereafter all of them will be named ``ions''), Birks proposed
semi-empirical formula \cite{Bir64,Bir51}:

\begin{equation}
\frac{dL}{dr} = \frac{S\frac{dE}{dr}}{1+kB\frac{dE}{dr}},
\end{equation}

\noindent where $BdE/dr$ is density of excitation centers along
the track, and $k$ is a quenching factor; $kB$ is usually treated
as a single parameter (Birks factor).

Equation (2) gives the following approximations for light yields for particles
with low (fast electrons) and high (ions) stopping power:

\begin{equation}
L_e(E) = SE, ~~~
L_i(E) = \frac{Sr}{kB},
\end{equation}

\noindent but in general light yield is:

\begin{equation}
L(E) = \int_0^E dL = \int_0^E \frac{SdE}{1+kB\frac{dE}{dr}}.
\end{equation}

Quenching factor for ions\footnote{Quenching factor for
$\alpha$ particles is often named ``$\alpha / \beta$'' ratio.} is a ratio of
light yield of ions to that of electrons of the same energy:

\begin{equation}
Q_i(E) =
\frac{L_i(E)}{L_e(E)} =
\frac{\int_0^E \frac{dE}{1+kB(\frac{dE}{dr})_i}}
     {\int_0^E \frac{dE}{1+kB(\frac{dE}{dr})_e}}.
\end{equation}

The $S$ factor disappeared in the ratio\footnote{Thus, the $S$ factor is supposed
independent on energy and equal for electrons and ions;
in the following we will suppose that the $kB$ factor is also independent on energy.}, 
and $Q_i(E)$ depends only on single parameter $kB$. 

Instead of quenching factor, sometimes a
relative light yield, ratio of $L_i$ to energy $E$, normalized to that for electron
$L_e$ at some energy $E_0$, is used:

\begin{equation}
R_i(E) = \frac{L_i(E)/E}{L_e(E_0)/E_0}.
\end{equation}

Relation between $Q_i(E)$ and $R_i(E)$ is evident:

\begin{equation}
R_i(E) =
\frac{L_i(E)}{L_e(E)}\frac{L_e(E)/E}{L_e(E_0)/E_0} =
Q_i(E)\frac{L_e(E)/E}{L_e(E_0)/E_0}.
\end{equation}

Thus, $R_i(E)$ is close to $Q_i(E)$ if $E$ and $E_0$ are in energy region where 
electron light yield $L_e$ is proportional to energy.

Taking into account that approximately (see Eq. (3)) $dL_e/dE = S$ and
$dL_i/dE = \frac{S}{kB} \frac{1}{(dE/dr)_i}$, we can obtain the following
approximation for quenching factor:

\begin{equation}
Q_i(E) =
\frac{L_i(E)}{L_e(E)} =
\frac{L_i(E)/E}{L_e(E)/E} \simeq
\frac{dL_i/dE}{dL_e/dE} \simeq
\frac{1}{kB(dE/dr)_i}.
\end{equation}

While this expression\footnote{In \cite{Bir64} it was erroneously written as
$Q_i = (dE/dr)_i/kB$, see Eq. (6.5).} is approximate, it gives the following
important features of quenching factor:

(1) Quenching factor depends on energy. This is not so trivial feature because
in many papers on $Q_i$ measurements it was supposed that $Q_i$ is constant;

(2) $Q_i$ is minimal when $(dE/dr)_i$ is maximal;

(3) $Q_i$ increases at low energies; this is a consequence of decrease of $(dE/dr)_i$,
see Fig. 1.

In the following, we will use Eq. (5) to calculate quenching
factors for different particles and different scintillators.
Stopping powers for ions will be calculated with the SRIM code
\cite{SRIM}. It should be noted that it is possible to calculate
stopping powers for $\alpha$ particles and protons also with the
ASTAR and PSTAR codes of the STAR package \cite{ESTAR},
respectively, but list of materials available is restricted. There
is also difference in $dE/dr$ calculated with the SRIM and ASTAR
\& PSTAR for $\alpha$ particles and protons which results also in
difference in calculated quenching factors; some examples are given later. The SRIM
code does not allow to calculate stopping powers for electrons, and
for this the ESTAR code \cite{ESTAR} will be used. Contrary to
other approaches, we will use {\em total} stopping powers instead
of using only electronic part of $dE/dr$. Currently we accept this
as a hypothesis and will show in future that it works well (in
particular, see Fig. 9 later).

Before to calculate quenching factors and compare them with experimental values
measured in different works, the following general note should be made.
Quenching factors could depend on many conditions of experimental measurements:

(1) If scintillator is not pure but doped with some material which enhances its
scintillating characteristics (e.g. Tl in NaI(Tl) or CsI(Tl)), $Q_i$ depends on
kind and amount of dopant. For example, for PbWO$_4$ detectors and external
$\alpha$ particles of 5.25 MeV $Q_i$ were measured as $Q_i = 0.19 - 0.32$
with different dopants and their different amounts \cite{Bar08}. And even if
some material is considered as a ``pure'' scintillator, usually it also contains
impurities and defects which could affect $Q_i$.

(2) Scintillation dynamics and light output depend on temperature. This is also
well known experimental fact; f.e. we can quote Ref. \cite{Ann08} where average decay
time of CaMoO$_4$ scintillator was measured as $\simeq17$ $\mu$s at $+20^\circ$ C but
as $\simeq350$ $\mu$s at $-140^\circ$ C. Usually scintillation has few components
with amplitudes different for different particles. Temperature dependence of these
amplitudes could lead to change in quenching factors;
for example, for $\alpha$ particle with $E_\alpha=2.14$ MeV emitted in $\alpha$
decay of $^{152}$Gd inside Gd$_2$SiO$_5$(Ce) detector, the $\alpha / \beta$ ratio
changed from 0.168 at $-20^\circ$ C to 0.178 at $+20^\circ$ C \cite{Yan07}.
In Ref. \cite{Sys98}, change in temperature from $+20^\circ$ C to $-20^\circ$ C
resulted in increase of the $\alpha / \beta$ ratio on 7\% in NaI(Tl), 35\% in CsI(Tl)
and 25\% in CsI(Na); change from $+20^\circ$ C 
to $+80^\circ$ C decreased $\alpha / \beta$ ratio
on 3\%, 15\% and 30\% in these crystals, respectively.

(3) Such a technical parameter as time $\Delta t$ during which scintillation signal
is collected by a data acquisition system, is in fact very important and could
drastically change $Q_i$ values. This is because the amplitudes of different
components of scintillation signal are different for different particles;
thus different parts of a signal will be collected during $\Delta t$.
This is illustrated by Fig. 2 where relative light output for CsI(Tl) and
external $\alpha$
particles with energies between $\simeq0$ and 10 MeV were measured with
$\Delta t = 1$ $\mu$s and 7 $\mu$s \cite{Gwi63}. Attempts to fit these data also are shown.
Descriptions of the data with Eq. (6) (which is better for $\Delta t = 7$ $\mu$s) lead
to different values of the $kB$ parameter:
$kB=1.1\times10^{-3}$ g MeV$^{-1}$ cm$^{-2}$ for $\Delta t = 1$ $\mu$s, and
$kB=2.3\times10^{-3}$ g MeV$^{-1}$ cm$^{-2}$ for $\Delta t = 7$ $\mu$s.
This gives very important conclusion: the Birks factor $kB$, which very
often is named ``Birks constant'', in fact {\em is not a fundamental constant}
for a given material but could have different values at different experimental
conditions (including time of a signal collection $\Delta t$).

\nopagebreak
\begin{figure}[htb]
\begin{center}
\mbox{\epsfig{figure=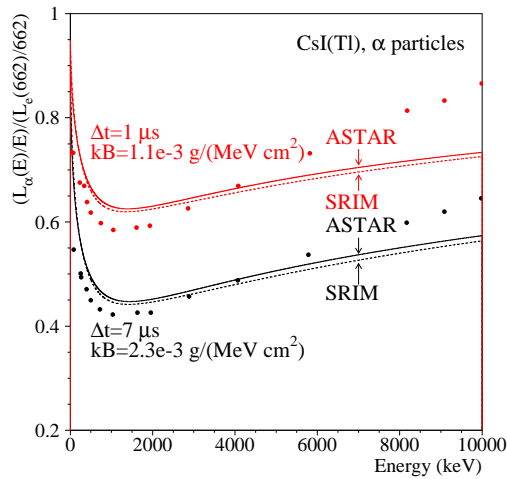,width=6.6cm}} 
\caption{(Color online) Relative
light output (normalized to that for electrons of 662 keV) for
CsI(Tl) and $\alpha$ particles \cite{Gwi63}. Fits of the data with
Eq. (6) are shown as continuous curves calculated with the ESTAR
(for electrons), and ASTAR or SRIM (for $\alpha$'s) codes.}
\end{center}
\end{figure}

If signals are not collected during a proper time, it is possible to obtain
wrong conclusions on $Q_i$ values. For example, the light output for protons
of 662 keV in CsI(Tl) measured during $\Delta t = 1$ $\mu$s in \cite{Gwi63}
is higher than that for $\gamma$ quanta of the same energy because scintillation 
signal for $p$ is faster. Thus, instead of expected quenching ($Q_i<1$) we, 
on contrary, obtain enhancement ($Q_i>1$). However, with $\gamma$ and $p$ signals 
collected during longer time of 7 $\mu$s, we have the usual situation when
$Q_i<1$ \cite{Gwi63}.

(4) Sometimes quenching factors are derived from measurements with non-monoenergetic
neutron sources like Am-Be\footnote{Spectrum of neutrons from Am-Be (Am-B) source
has complex structure with energies up to $\simeq11$ MeV ($\simeq5.5$ MeV) \cite{Mar95}.}.
After collision with neutron, nuclear recoil with mass $M_r$ has energy determined
by initial energy of neutron $E_n$ and neutron scattering angle $\theta$:

\begin{equation}
E_r = \frac{2E_n}{(1+\mu)^2} (\mu + \sin^2 \theta -\cos \theta \sqrt{\mu^2 - \sin^2 \theta})
\simeq \frac{2E_n}{\mu} (1 - \cos \theta),
\end{equation}

\noindent where $\mu = M_r/m_n$, $m_n$ is the neutron mass, and the last approximation
is valid for heavy nuclei ($\mu \gg 1$). Energy dependence of quenching factor can
be found only if the initial neutron energy and scattering angle are known;
measurements with non-monoenergetic neutron sources give only some average
$Q_i$ value which, nevertheless, could be useful estimate of $Q_i$ for energy
range $\simeq 0 - 4E_n/A$, where $A$ is mass number of nuclear recoil,
if more detailed data are absent (this effective energy range is also related 
with dependence of the cross section on angle of scattering).

(5) Presence of electric field in case of liquid noble gases could distort
initial $Q_i$ values (obtained without electric field).
Other phenomena like channeling effect in crystals (see f.e. \cite{Ber08c}),
dependence of $Q_i$ on direction of particle's movement relatively to
crystal axes (as f.e. in CdWO$_4$ \cite{Dan03}) or diffusion and movement
of molecules in liquids \cite{Anr09} also effect quenching factors.

In accordance with the above mentioned, we will not expect that the Birks factor
$kB$ for a given material will have the same value in different measurements.
However, we will expect that if conditions of measurements and data treatment
are fixed, $kB$ will be the same for all particles.
Such a hypothesis was discussed already in \cite{Bir64}, and it was supported
by some experimental data. Below we show that it
gives reliable results for a range of energies of interest here (low energy
ions and $\alpha$ particles with energies up to $\simeq10$ MeV).

\section{Calculation of quenching factors}

Results of calculation of quenching factors with Eq. (5), or
relative light outputs with Eq. (6) are presented below for a
number of organic, crystal and liquid noble gases scintillators. To compare
calculations with experimental results, among big number of
experimental papers we mainly chose more recent articles where
data were obtained with better techniques (monoenergetic
neutron beam instead of Am-Be source, for example) or/and in wider
energy range.

\subsection{Organic scintillators}

\subsubsection{Polystyrene (C$_8$H$_8$)}

Relative light output $L_\alpha(E)/E$ for $\alpha$ particles with
energies of $2-9$ MeV (normalized to that for electrons of 976
keV\footnote{Difference between the relative light output and
quenching factor for presented energies is small.}) for
polystyrene scintillator (chemical formula C$_8$H$_8$, density
$\rho=1.06$ g cm$^{-3}$) was measured recently in Ref.
\cite{Bon07} using external $\alpha$ particles from $^{241}$Am
source with a set of thin mylar absorbers, and 
internal $\alpha$ particles from contamination of the scintillator
by U/Th chains. Fit of these
experimental data by calculations with Eq. (6) and with the Birks
factor $kB=9.0\times10^{-3}$ g MeV$^{-1}$ cm$^{-2}$ is presented
in Fig. 3a. Fitting curves were obtained with stopping powers
calculated with the SRIM or ASTAR codes for $\alpha$ particles,
and ESTAR code for electrons. Difference between the ASTAR and SRIM 
calculations is not very big, 
however $\chi^2$/n.d.f. value\footnote{Values of $\chi^2$/n.d.f. are calculated
everywhere without taking into account uncertainties in energy.} 
for the ASTAR (0.50) is better than that for the SRIM (0.58).

\nopagebreak
\begin{figure}[htb]
\begin{center}
\mbox{\epsfig{figure=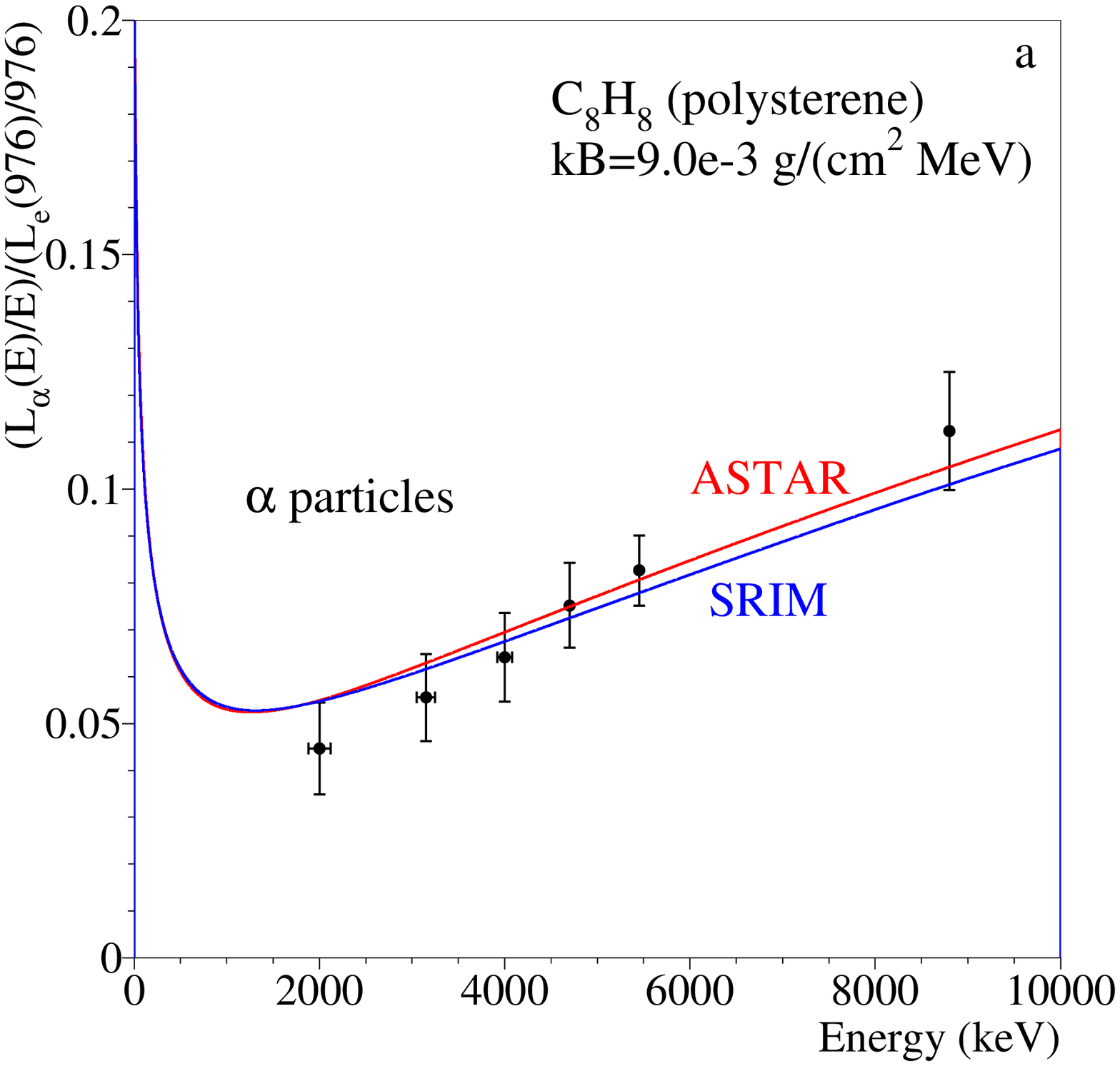,width=6.6cm}}~
\mbox{\epsfig{figure=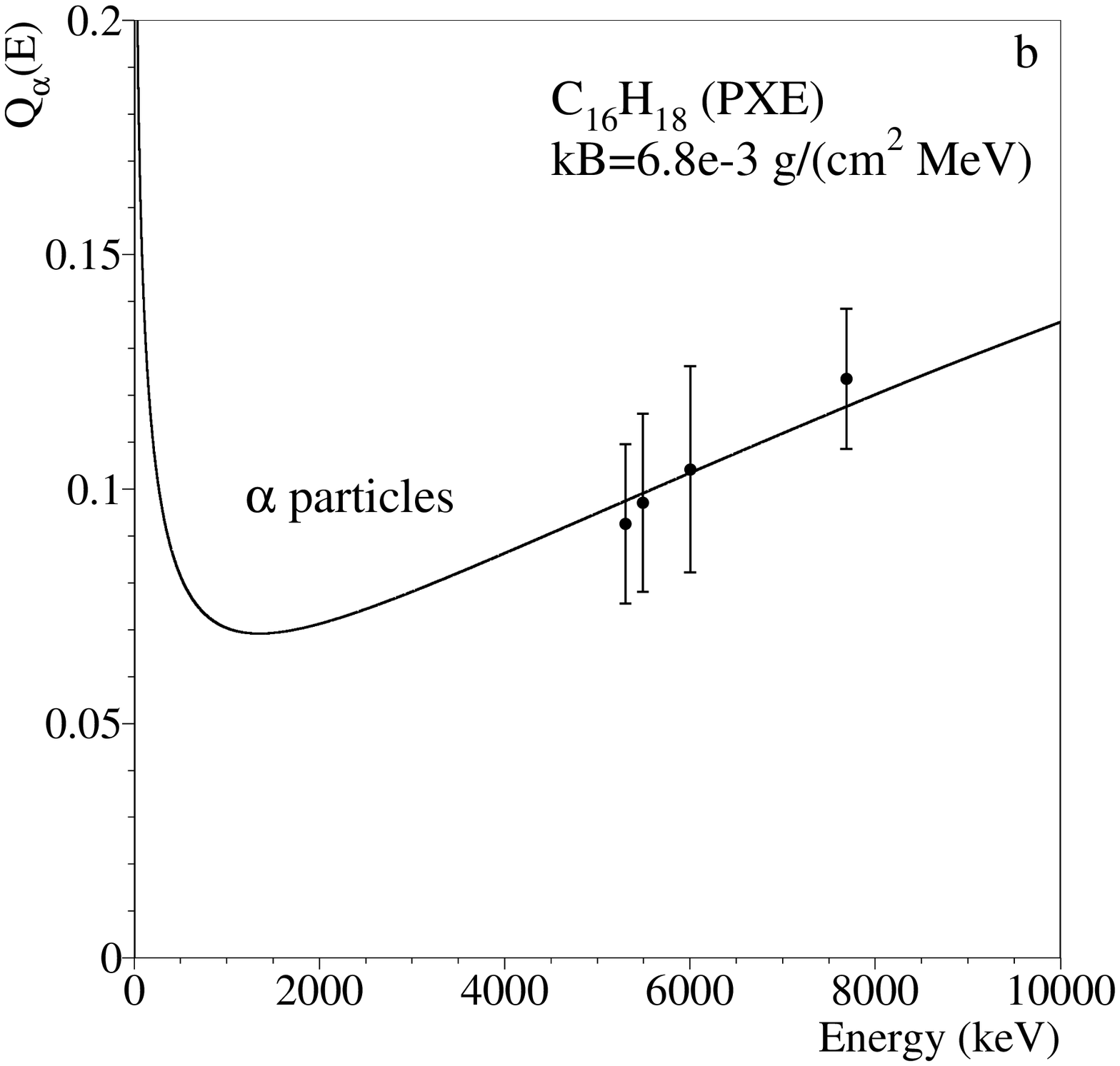,width=6.6cm}} \\
\mbox{\epsfig{figure=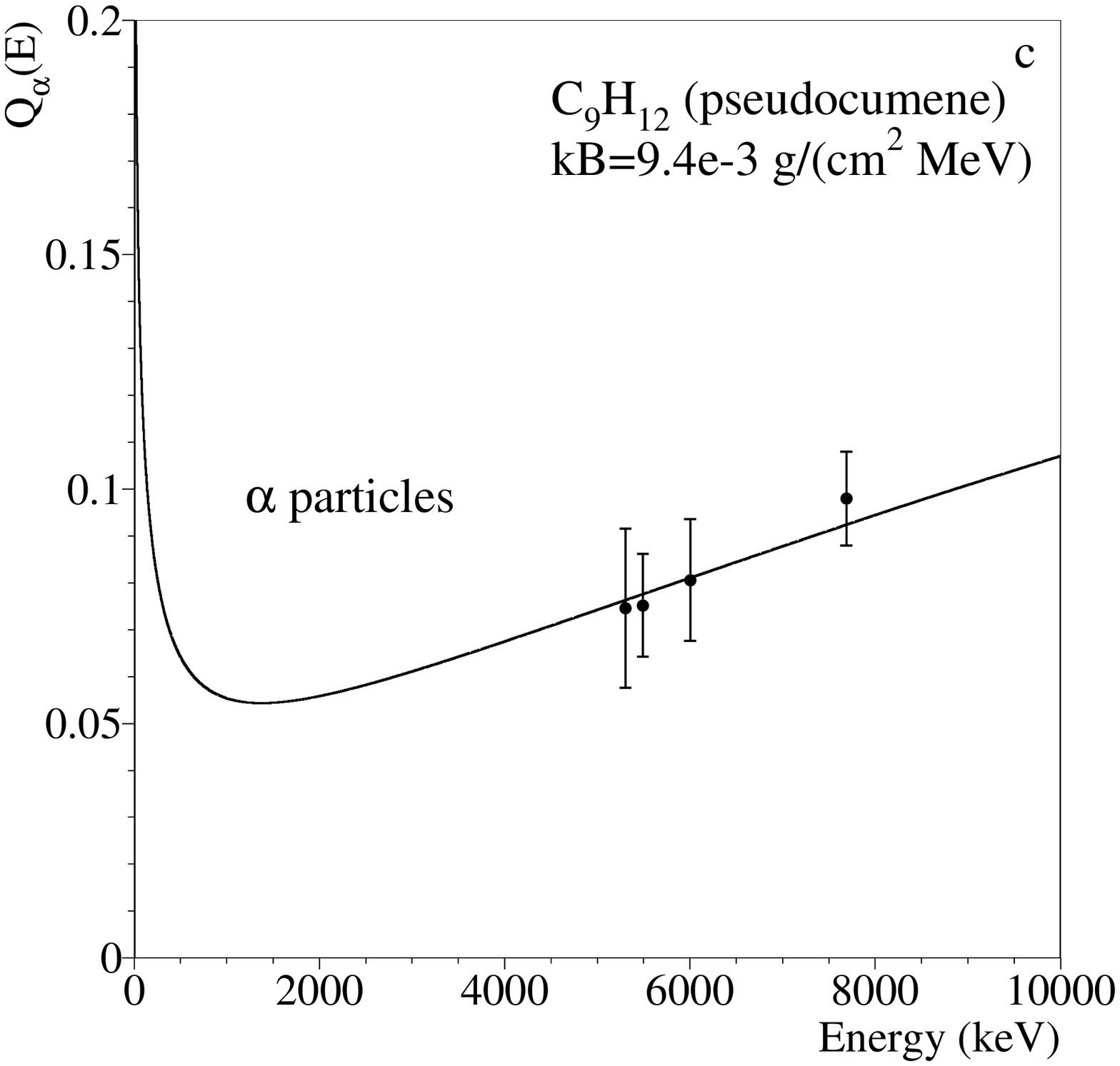,width=6.6cm}}~
\mbox{\epsfig{figure=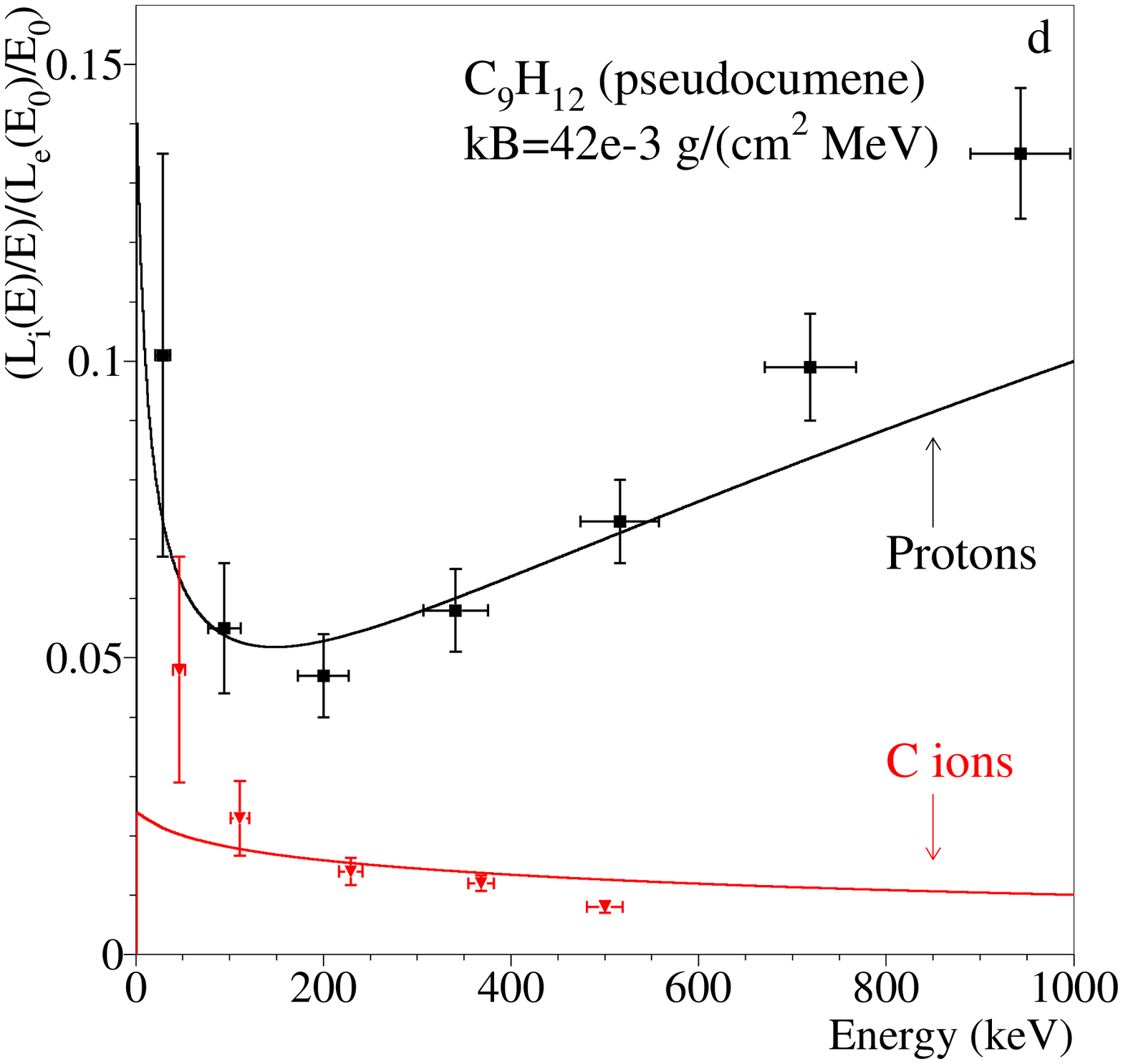,width=6.6cm}}
\caption{(Color online) Quenching factors or relative light outputs
for $\alpha$ particles and ions in organic scintillators:
experimental data and their fit with Eq. (5) or (6).
(a) C$_8$H$_8$ scintillator (polystyrene) \cite{Bon07},
$kB=9.0\times10^{-3}$ g MeV$^{-1}$ cm$^{-2}$;
(b) C$_{16}$H$_{18}$ liquid scintillator (PXE) \cite{Bac08},
$kB=6.8\times10^{-3}$ g MeV$^{-1}$ cm$^{-2}$;
(c) C$_{9}$H$_{12}$ liquid scintillator (pseudocumene) \cite{Bac08},
$kB=9.4\times10^{-3}$ g MeV$^{-1}$ cm$^{-2}$;
(d) C$_{9}$H$_{12}$ \cite{Hon02},
$kB=35\times10^{-3}$ g MeV$^{-1}$ cm$^{-2}$;
$E_0=22$ keV for protons and 32 keV for C ions.
In part (a), results with stopping powers calculated with the ASTAR and SRIM codes
for $\alpha$ particles are shown.}
\end{center}
\end{figure}

\subsubsection{PXE (C$_{16}$H$_{18}$)}

In Ref. \cite{Bac08}, quenching factors for $\alpha$ particles were investigated
for liquid scintillator:
phenyl-o-xylylethane (1,2-dimethyl-4-(1-phenylethyl)-benzene, PXE,
chemical formula C$_{16}$H$_{18}$, density $\rho=0.988$ g cm$^{-3}$) doped with
para-Terphenyl (1,4-diphenylbenzene, p-Tp) at 2.0 g/l and
bis-MSB (1,4-bis(2-methylstyryl)\-benzene) at 20 mg/l.
Alpha particles belong to internal contamination of scintillator by daughters
from $^{238}$U chain ($^{222}$Rn, $^{218}$Po, $^{214}$Po, $^{210}$Po);
$E_\alpha=5.3-7.7$ MeV.

In fact, in case of $\alpha$ decay of nucleus inside a scintillator,
released light output has two components:
from $\alpha$ particle and from nuclear recoil.
However, corrections for the light output from heavy nuclear recoils are small:
for example, for $\alpha$ decay of $^{210}$Po ($Q_\alpha=5407$ keV) corresponding
energies are $E_\alpha=5304$ keV, $E_r=103$ keV and with the value of
$kB=6.8\times10^{-3}$ g MeV$^{-1}$ cm$^{-2}$ (see Fig. 3b)
the light output from $^{206}$Pb recoil is only 2.1\% of that from the $\alpha$ particle.
We will neglect these corrections in the following.
Quenching factors calculated for
C$_{16}$H$_{18}$ (PXE) with $kB=6.8\times10^{-3}$ g MeV$^{-1}$ cm$^{-2}$
are shown in Fig. 3b in comparison with the experimental data \cite{Bac08}
($\chi^2$/n.d.f. = 0.08).

\subsubsection{Pseudocumene (C$_{9}$H$_{12}$)}

(1) In Ref. \cite{Bac08}, quenching factors for $\alpha$ particles
were studied as above also for another liquid scintillator:
pseudocumene (1,2,4-trimethylbenzene, PC, C$_9$H$_{12}$,
$\rho=0.876$ g cm$^{-3}$) doped with 2,5-diphenyloxazole (PPO) at
a concentration of 1.5 g/l. Fit of these data ($\chi^2$/n.d.f. = 0.12)
is shown in Fig. 3c
with $kB=9.4\times10^{-3}$ g MeV$^{-1}$ cm$^{-2}$.

(2) Relative light outputs $L_i(E)/E$ for protons and C ions in
pseudocumene (BC505 liquid scintillator) were measured in
\cite{Hon02} in range of energies of $29-943$ keV (protons) and
$46-500$ keV (C ions). They were normalized to relative light
output for electrons at 22 keV (for $p$) and 32 keV (for C)
\cite{Hon02}. The Birks factor $kB=42\times10^{-3}$ g MeV$^{-1}$
cm$^{-2}$ was found by fitting the experimental data for protons
by Eq. (6) (see Fig. 3d); $\chi^2$/n.d.f. value is 2.8, with main 
contribution due to the last experimental point. 
After this, the curve for C ions was calculated with this $kB$ 
value. Comparison of the curve with the C data gives
$\chi^2$/n.d.f. = 5.3. While this value is high, nevertheless 
calculations are in a proper agreement with the measured data, as one can
see in Fig. 3d.

The $kB$ values for pseudocumene C$_9$H$_{12}$ obtained by fitting data of
\cite{Bac08} and \cite{Hon02}, respectively (Fig. 3c and 3d),
are quite different. However, this is not a surprise taking into account different
conditions of measurements and probably dopants used. Nevertheless,
the same $kB$ value fixed in one experiment \cite{Hon02} allowed to describe
both data sets: for protons and C ions.

\subsection{Crystal scintillators}

\subsubsection{CdWO$_4$}

(1) Quenching factors for $\alpha$ particles in CdWO$_4$ crystal scintillator
(density $\rho=7.9$ g cm$^{-3}$) were
measured in experimental searches for rare $\alpha$ decay of $^{180}$W
($Q_\alpha=2516$ keV) with CdWO$_4$ detector \cite{Dan03} where it was observed
at the first time ($T_{1/2}=1.1\times10^{18}$ yr).
Alpha particles from external $^{241}$Am source (with a set of thin 
absorbers\footnote{It should be noted that quenching factors in CdWO$_4$
depend on direction of movement of $\alpha$ particle;
we use here data for direction perpendicular to (010) crystal plane
measured in wider energy range \cite{Dan03}.}) 
and from internal contamination of CdWO$_4$ by U/Th chains were used;
$E_\alpha=0.47-8.79$ MeV.
The experimental points and fitting curve calculated with Eq. (5) and
$kB=10.1\times10^{-3}$ g MeV$^{-1}$ cm$^{-2}$ are shown in Fig. 4a
($\chi^2$/n.d.f. = 13; without last point $\chi^2$/n.d.f. = 2.2).

\nopagebreak
\begin{figure}[htb]
\begin{center}
\mbox{\epsfig{figure=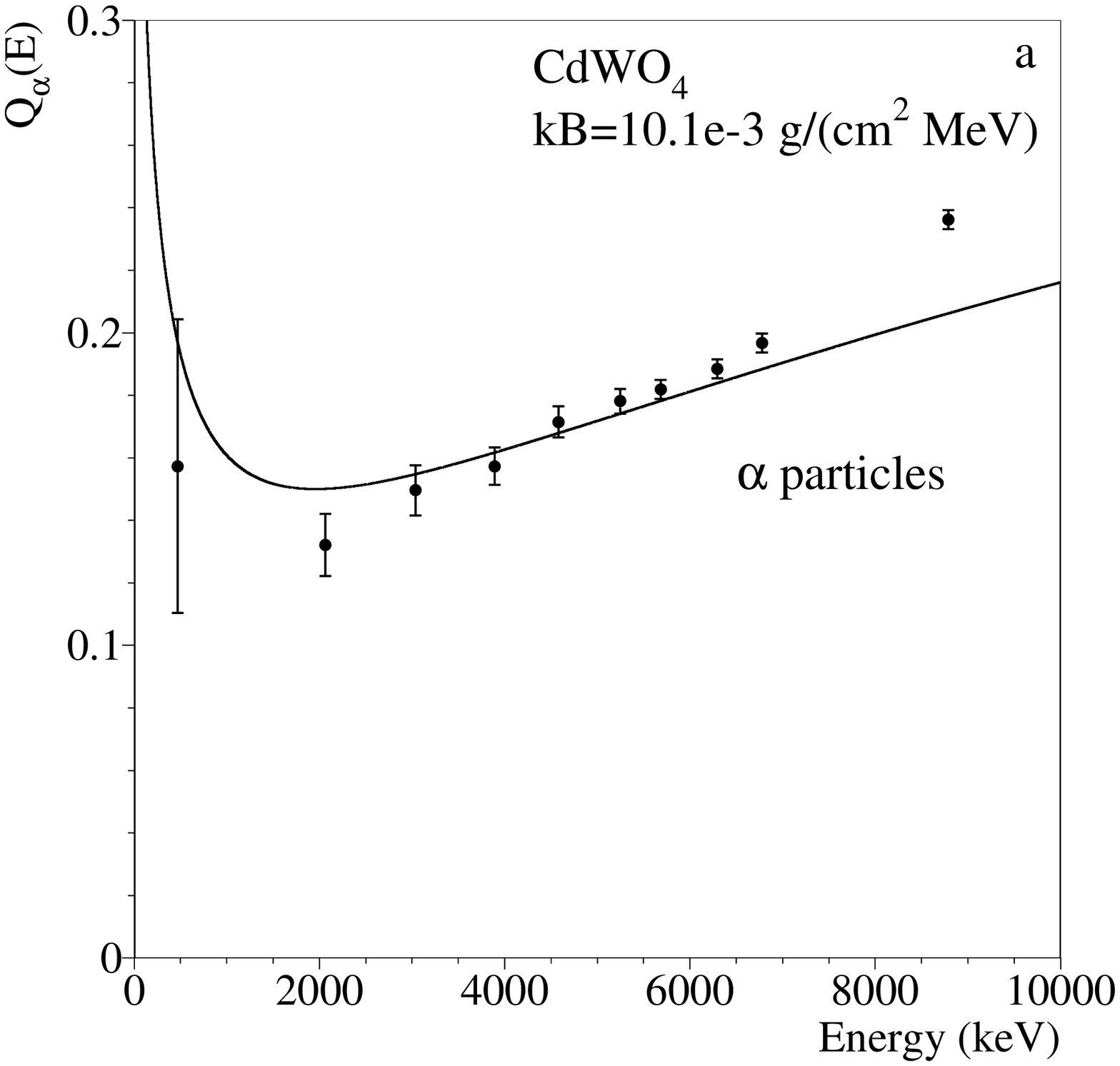,width=6.6cm}}~
\mbox{\epsfig{figure=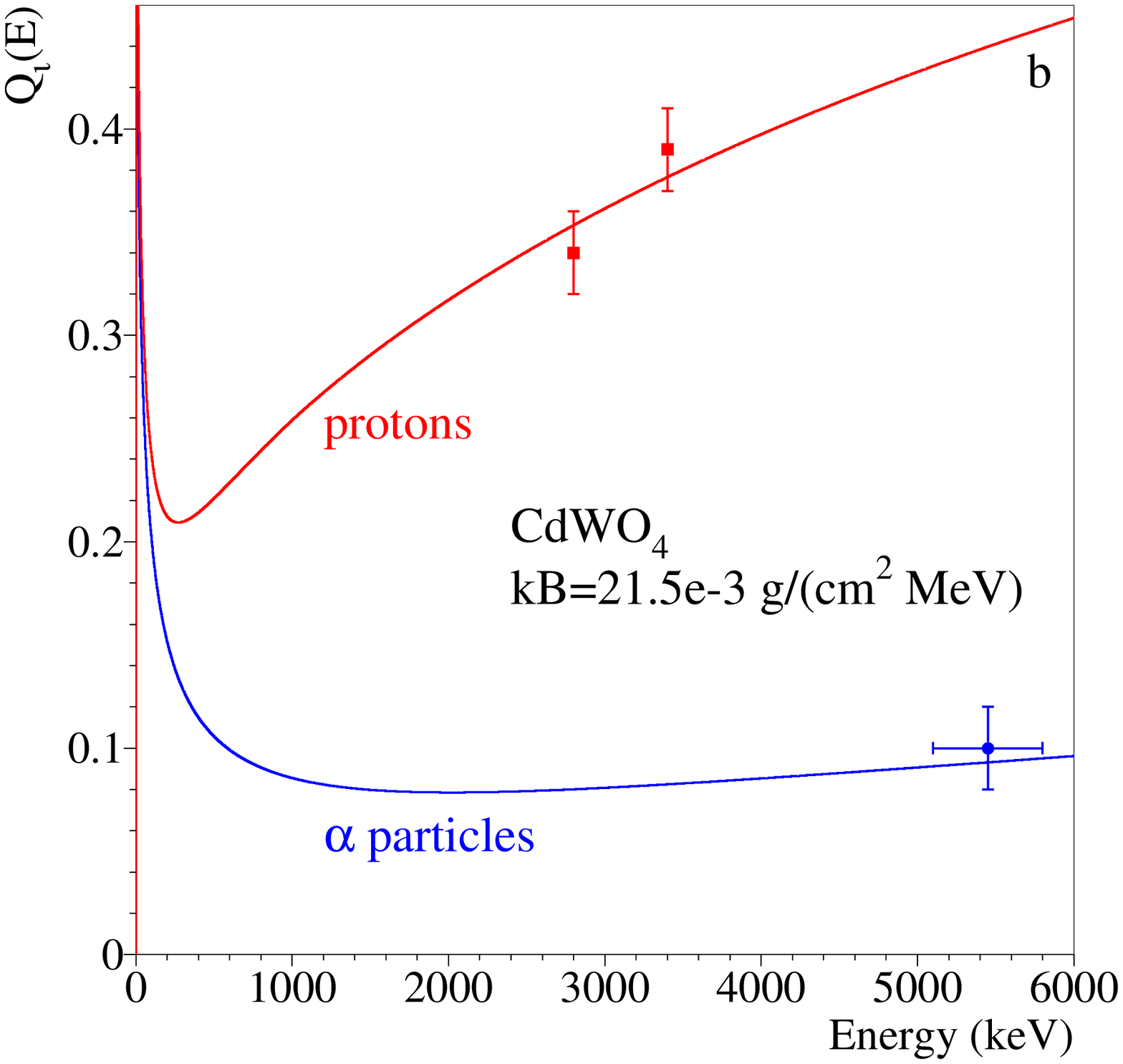,width=6.6cm}}
\caption{(Color online) Quenching factors for $\alpha$ particles and protons in CdWO$_4$.
(a) Experimental data from Ref. \cite{Dan03} and their fit by Eq. (5)
with $kB=10.1\times10^{-3}$ g MeV$^{-1}$ cm$^{-2}$;
(b) data for protons and $\alpha$'s from \cite{Faz98} and their fit
with $kB=21.5\times10^{-3}$ g MeV$^{-1}$ cm$^{-2}$.}
\end{center}
\end{figure}

(2) In Ref. \cite{Faz98}, quenching factors for protons provided by accelerator with
$E_p=2.8$ MeV and 3.4 MeV were measured as $0.34\pm0.02$ and $0.39\pm0.02$,
respectively. Fit of these points by Eq. (5) ($\chi^2$/n.d.f. = 0.90)
was possible with the value of the
Birks factor: $kB=21.5\times10^{-3}$ g MeV$^{-1}$ cm$^{-2}$,
very different from that obtained for CdWO$_4$ by fitting data from Ref. \cite{Dan03}.
Quenching factors for $\alpha$ particles were also measured in \cite{Faz98}.
A mixed source with $\alpha$'s from $^{239}$Pu ($E_\alpha \simeq 5.1$ MeV \cite{ToI98}),
$^{241}$Am ($E_\alpha \simeq 5.5$ MeV) and $^{244}$Cm ($E_\alpha \simeq 5.8$ MeV) was 
used. Quenching factor of $Q_\alpha\simeq0.1$ was measured (different from
those in \cite{Dan03}).
The value of $kB=21.5\times10^{-3}$ g MeV$^{-1}$ cm$^{-2}$ obtained for protons
gives theoretical curve for $\alpha$'s in good agreement with this experimental result
($\chi^2$/n.d.f. = 0.12), see Fig. 4b.

\subsubsection{CaF$_2$(Eu)}

(1) Quenching factors for $\alpha$ particles in CaF$_2$ crystal scintillator
($\rho=3.18$ g cm$^{-3}$) doped by Eu at 0.4\% were measured in \cite{Bel07}
with external ($^{241}$Am) and internal (U/Th chains, $^{147}$Sm) $\alpha$ sources
($E_\alpha=1-9$ MeV). The measured experimental points and their fit by Eq. (5)
with the Birks factor $kB=5.3\times10^{-3}$ g MeV$^{-1}$ cm$^{-2}$ are presented
in Fig. 5a. Stopping powers of $\alpha$ particles were calculated with the SRIM
and ASTAR codes; fitting with the ASTAR curve is better:
$\chi^2$/n.d.f. is 3.5 (with near 50\% contribution from the last point) while
for the SRIM $\chi^2$/n.d.f. = 7.6.
It is a pity that the STAR package allows calculations of $dE/dr$ only for restricted
list of materials in case of $\alpha$ particles and protons (and other ions are absent).
From the other side, SRIM allows to calculate $dE/dr$ for any materials and for any ions,
but not for electrons. Fig. 5a gives an idea that probably calculations of $dE/dr$
for all particles inside the same package would give better description of
quenching factors.

\nopagebreak
\begin{figure}[htb]
\begin{center}
\mbox{\epsfig{figure=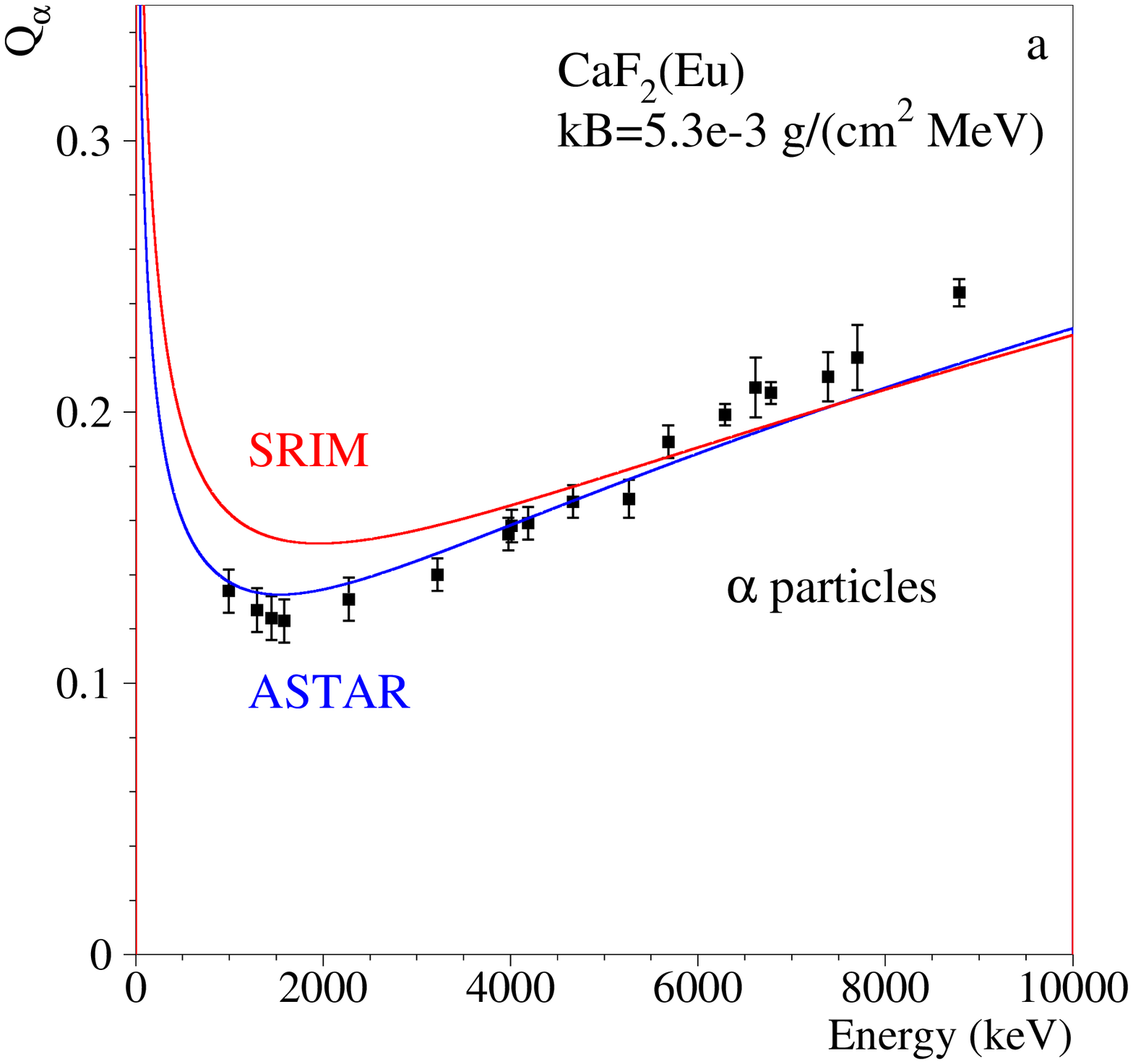,width=6.6cm}}~
\mbox{\epsfig{figure=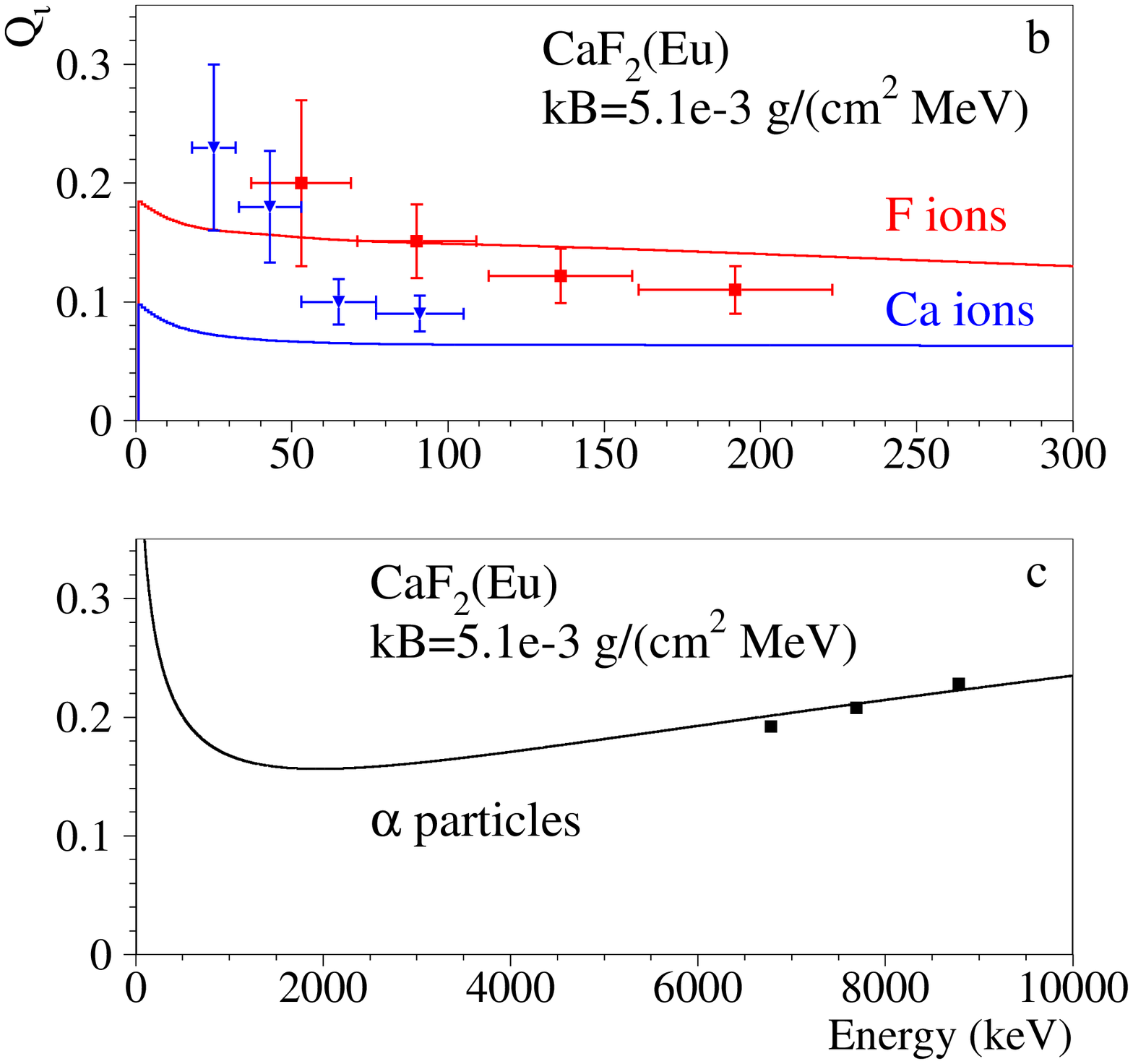,width=6.6cm}}
\caption{(Color online) 
Quenching factors for $\alpha$ particles, and F and Ca ions in CaF$_2$(Eu).
(a) Experimental data from Ref. \cite{Bel07} for $\alpha$ particles
and their fit by Eq. (5)
with $kB=5.3\times10^{-3}$ g MeV$^{-1}$ cm$^{-2}$ with stopping powers
calculated with the SRIM and ASTAR.
(b) Data for F (squares) and Ca (triangles) ions from \cite{Haz02} and their fit
with $kB=5.1\times10^{-3}$ g MeV$^{-1}$ cm$^{-2}$.
(c) Data from \cite{Ume08} for $\alpha$'s with the same $kB$ value as in (b).}
\end{center}
\end{figure}

(2) Quenching factors for F and Ca ions in CaF$_2$(Eu) measured in \cite{Haz02}
are shown in Fig. 5b. The $kB$ value was obtained by fitting the F data as:
$kB=5.1\times10^{-3}$ g MeV$^{-1}$ cm$^{-2}$ ($\chi^2$/n.d.f. = 1.3).
Curve for Ca ions was calculated with this $kB$; 
however, agreement with experimental points is not so good
($\chi^2$/n.d.f. = 4.3).

(3) Quenching factors for $\alpha$ particles in CaF$_2$(Eu), obtained by the same group
as in Ref. \cite{Haz02},
can be derived from their paper \cite{Ume08}. Alpha peak from $^{216}$Po
($E_\alpha=6.778$ MeV \cite{ToI98}) was observed at energy of 1.3 MeV,
from $^{214}$Po ($E_\alpha=7.687$ MeV) -- at 1.6 MeV, and
from $^{212}$Po ($E_\alpha=8.784$ MeV) -- at 2.0 MeV;
thus corresponding quenching factors are equal 0.192, 0.208 and 0.228, respectively.
Because we could expect the same (or similar) conditions of measurements and data
treatment in both \cite{Haz02} and \cite{Ume08}, the value of
$kB=5.1\times10^{-3}$ g MeV$^{-1}$ cm$^{-2}$ could fit the data for $\alpha$'s
as well. Such a curve is presented in Fig. 5c and is in good agreement
with the experimental data.

\subsubsection{PbWO$_4$}

Quenching factors for $\alpha$ particles in PbWO$_4$ crystal scintillator
($\rho=8.28$ g cm$^{-3}$) were studied in \cite{Dan06} in the range of energies
$E_\alpha=2.1-5.3$ MeV. They are shown in Fig. 6 together with fit by Eq. (5) with
$kB=10.5\times10^{-3}$ g MeV$^{-1}$ cm$^{-2}$ ($\chi^2$/n.d.f. = 3.3).

\nopagebreak
\begin{figure}[htb]
\begin{center}
\mbox{\epsfig{figure=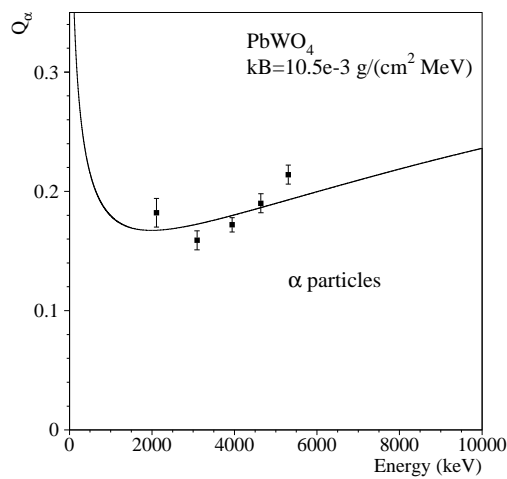,width=6.6cm}}
\caption{Quenching factors for $\alpha$ particles in PbWO$_4$ measured
in \cite{Dan06} and their fit with $kB=10.5\times10^{-3}$ g MeV$^{-1}$ cm$^{-2}$.}
\end{center}
\end{figure}

\subsubsection{ZnWO$_4$}

(1) ZnWO$_4$ ($\rho=7.8$ g cm$^{-3}$) is one of perspective scintillators in searches
for dark matter particles (see e.g. \cite{Bav08,Kra09}). Quenching factors for
$\alpha$ particles were studied in \cite{Dan05}\footnote{As in CdWO$_4$,
quenching factors in ZnWO$_4$ depend on direction of movement of $\alpha$ particle;
we use here data for direction perpendicular to (010) crystal plane
measured in wider energy range \cite{Dan05}.}. 
They are shown in Fig. 7a
together with fit by Eq. (5) with $kB=9.0\times10^{-3}$ g MeV$^{-1}$ cm$^{-2}$;
the agreement between the calculated curve and the experimental data is good
($\chi^2$/n.d.f. = 0.93).

\nopagebreak
\begin{figure}[htb]
\begin{center}
\mbox{\epsfig{figure=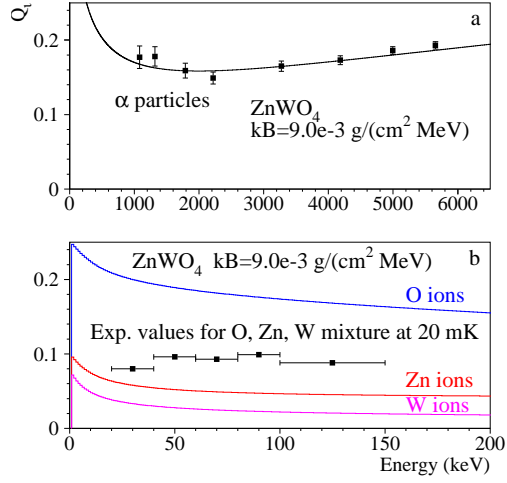,width=6.6cm}}
\caption{(Color online) 
(a) Quenching factors for $\alpha$ particles in ZnWO$_4$ measured in
\cite{Dan05} and their fit with the Birks factor
$kB=9.0\times10^{-3}$ g MeV$^{-1}$ cm$^{-2}$.
(b) Quenching factors for O, Zn and W ions in ZnWO$_4$ calculated
with this $kB$ value together with experimental data for mixture
of O, Zn, W factors measured at temperature $\simeq20$ mK \cite{Bav08}.}
\end{center}
\end{figure}

(2) In Fig. 7b, calculated quenching factors for O, Zn and W ions obtained with
this $kB$ value are presented. Also shown are experimental data collected
with ZnWO$_4$ detector in bolometric measurements \cite{Bav08}.
Comparing calculations with these data, we have to remember that:
(a) the latter are in fact data for some mixture of quenching factors for O, Zn
and W ions;
(b) predictions are given for $kB=9.0\times10^{-3}$ g MeV$^{-1}$ cm$^{-2}$
derived by fitting data for $\alpha$ particles measured at room temperature
while in Ref. \cite{Bav08} temperature was $\simeq20$ mK.
To obtain reliable predictions, data for all particles should be collected
at the same conditions (and with the same data treatment).

\subsubsection{CaWO$_4$}

Quenching factors for CaWO$_4$ scintillator ($\rho=6.06$ g cm$^{-3}$)
probably are the most extensively investigated, in particular, because of
numerous studies in the CRESST experimental searches for dark matter \cite{Ang05}.

(1) Values for $\alpha$ particles were measured with external and internal $\alpha$
sources in \cite{Zde05} in the energy range $E_\alpha \simeq 0.5-8$ MeV.
These data are shown in Fig. 8 together with their fit by Eq. (5) with
$kB=6.2\times10^{-3}$ g MeV$^{-1}$ cm$^{-2}$
($\chi^2$/n.d.f. = 6.4).

\nopagebreak
\begin{figure}[htb]
\begin{center}
\mbox{\epsfig{figure=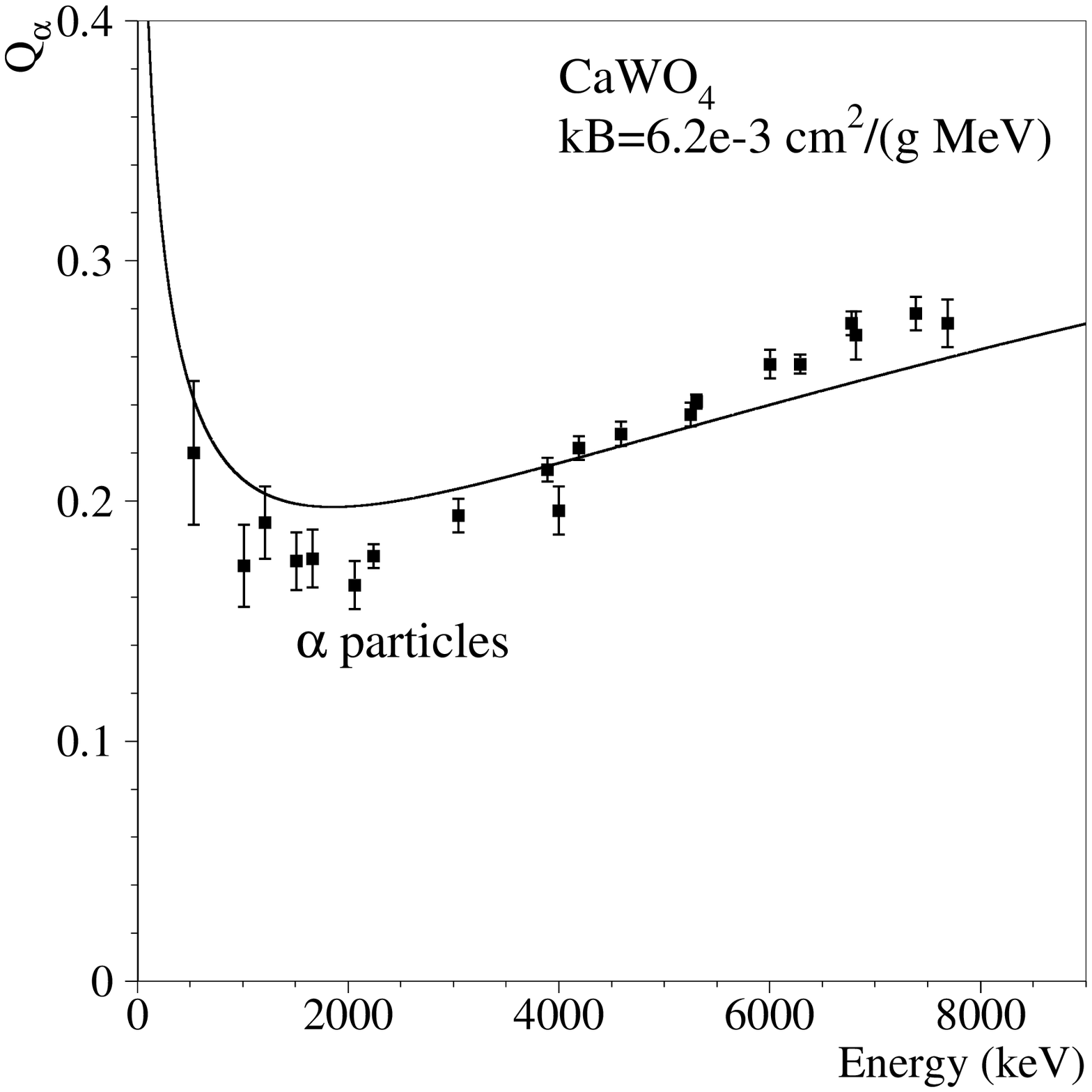,width=6.6cm}}
\caption{Quenching factors for $\alpha$ particles in CaWO$_4$ measured in
\cite{Zde05} and their fit with $kB=6.2\times10^{-3}$ g MeV$^{-1}$
cm$^{-2}$.}
\end{center}
\end{figure}

(2) Energy dependence of quenching factors for O, Ca and W ions in CaWO$_4$ was measured
with monoenergetic neutron beam in \cite{Jag06} (for W, only upper limits of $Q_i$
were obtained, see Fig. 9a). Data for O ions were fitted by Eq. (5); the obtained
value of the Birks factor is: $kB=8.0\times10^{-3}$ g MeV$^{-1}$ cm$^{-2}$
(no surprise that it is different from that determined above for $\alpha$ particles
due to different conditions of measurements and data treatment)\footnote{Formally,
$\chi^2$/n.d.f. value for O ions is very high (28.3) that is related with the small error bars
in the experimental data. However, it should be noted that deviations of the calculated
values from the experimental ones are in the range of only 2 -- 12\%.}.
Now, fixing this $kB$ value, we calculate quenching factors for Ca and W ions; all
results are shown in Fig. 9a. One can say that calculations for O ions are
in reliable agreement with the experimental points
($\chi^2$/n.d.f. = 4.8), and also curve for W ions is not
in contradiction with the measured W limits.

\nopagebreak
\begin{figure}[htb]
\begin{center}
\mbox{\epsfig{figure=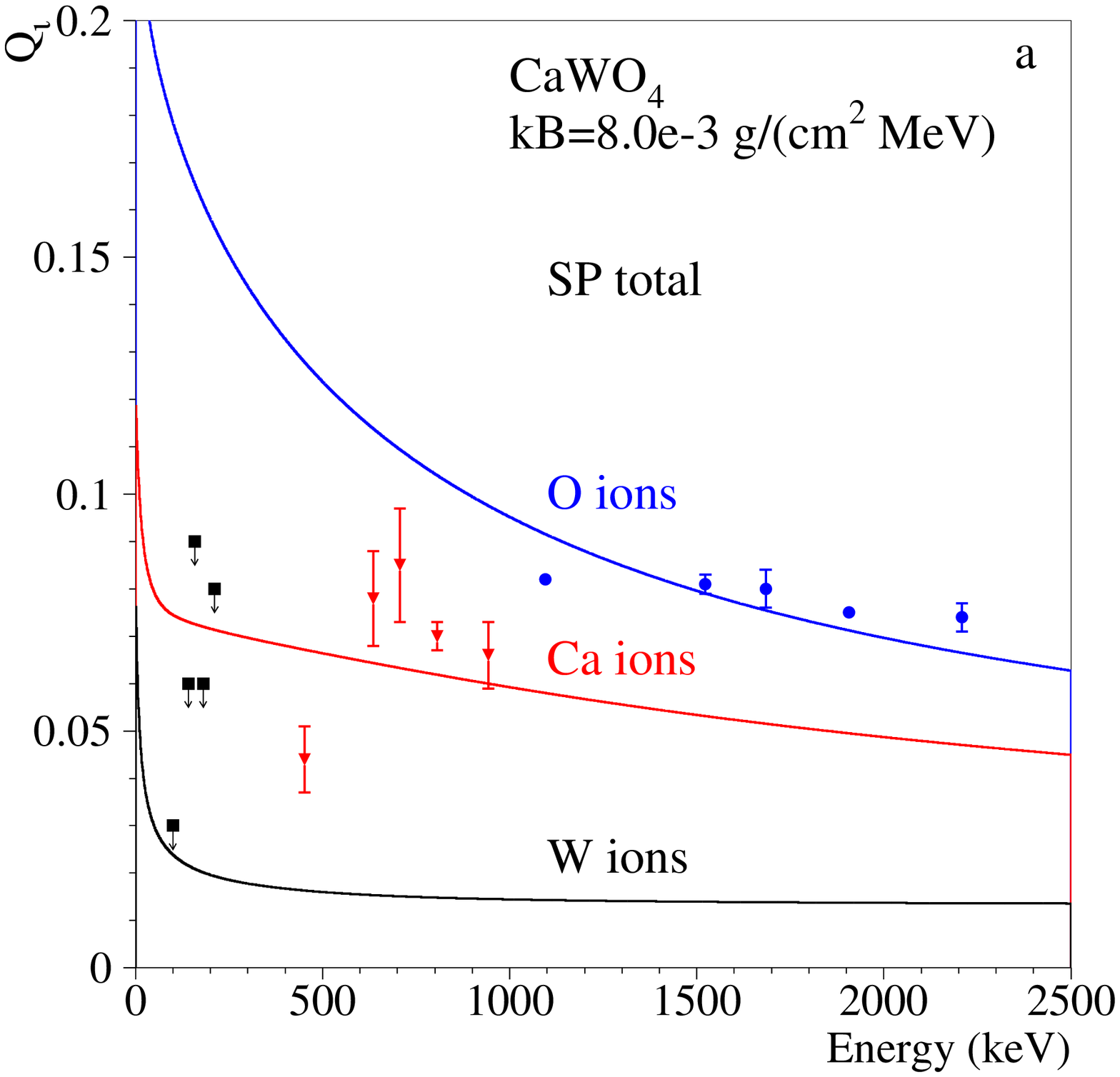,width=6.6cm}}~
\mbox{\epsfig{figure=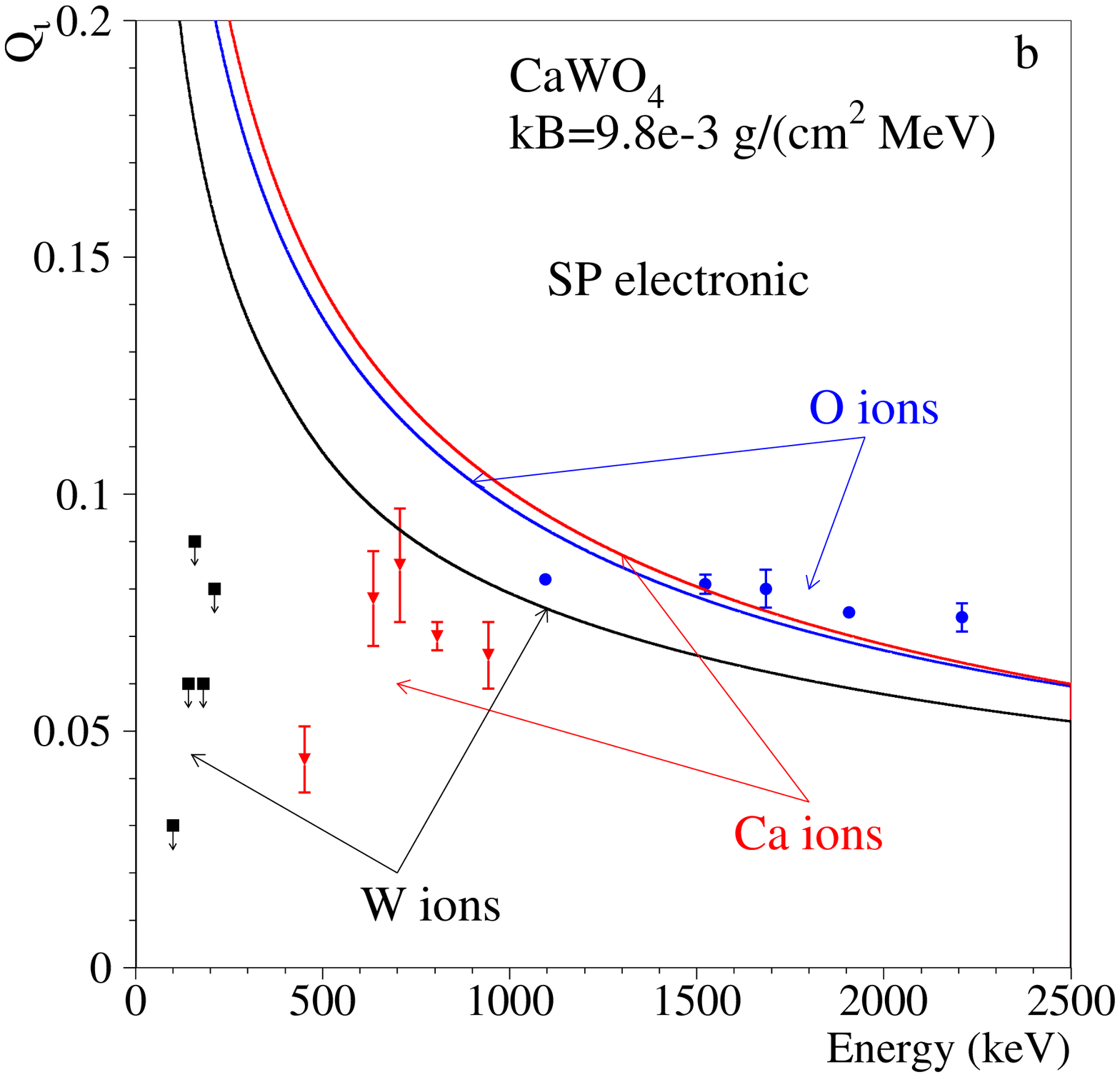,width=6.6cm}}
\caption{(Color online) 
Quenching factors for O (circles), Ca (triangles)
and W (squares) ions in CaWO$_4$ measured in \cite{Jag06} and
their fits:
(a) with the ion total SP and $kB=8.0\times10^{-3}$ g MeV$^{-1}$ cm$^{-2}$;
(b) with only electronic part of the SP 
($kB=9.8\times10^{-3}$ g MeV$^{-1}$ cm$^{-2}$).}
\end{center}
\end{figure}

(3) We want to return here to discussion on which stopping power (SP)
-- total or only electronic part --
is better to use in fitting experimental data. Using the same ideology,
the experimental results for O ions were fitted by Eq. (5) where only electronic
part of ion SP was taken into account. Obtained Birks factor was equal
$kB=9.8\times10^{-3}$ g MeV$^{-1}$ cm$^{-2}$; fit is worse than that with the total SP
(see Fig. 9b).
With this $kB$ value, quenching curves for Ca and W ions were calculated;
they are in evident disagreement with the experimental data.
Thus, use of the ion total SP in fitting and predicting quenching factors in the
proposed approach allows to describe experimental data in a much better way.

(4) Dependence of the light output and quenching factors in CaWO$_4$
on atomic mass of ion was measured in Ref. \cite{Nin06} by
impinging various ions -- from H to Au -- onto the scintillating
crystal. All ions had energy of 18 keV. Inverse values to the
relative light outputs of Eq. (6), normalized to that of electrons
at 6 keV, were calculated and presented in Table 4 of Ref.
\cite{Nin06}:

\begin{equation}
R'_i = \frac{1}{R_i(E)} = \frac{L_e(E_0)/E_0}{L_i(E)/E}
\end{equation}

with $E=18$ keV and $E_0=6$ keV; values of $R'_i$ changed from $\simeq2$ for H
to $\simeq40$ for Au. Remembering that the $kB$ values could be
different under different experimental conditions and data
treatment, we will use $R'_i$ for H to determine $kB$ in this
particular measurements and will use this $kB$ value to calculate
$R'_i$ for all other ions. Value of $kB=17\times10^{-3}$ g
MeV$^{-1}$ cm$^{-2}$ gives $R'_{\mathrm{H}}=2.17$ that well reproduces
values for H given in Table 4 of \cite{Nin06} ($2.15\pm0.02$ for
collection of the scintillation signal during $\Delta t = 40$
$\mu$s, and $2.18\pm0.02$ for $\Delta t = 50$ $\mu$s). $R'_i$ values
for other ions calculated with this $kB$ value are shown in Fig.
10a together with experimental results\footnote{We use data for $\Delta t = 50$
$\mu$s that gives more complete collection of scintillating signal;
however values for $\Delta t = 40$ $\mu$s are close.} of Ref.
\cite{Nin06} in dependence on ion's $Z$ number. Theoretical points
lay on smooth curve well fitted by polynomial $R'_i(Z) = a + bZ +
cZ^2$ with $a=0.87098$, $b=0.98708$ and $c=-5.9896\times10^{-3}$
(also shown in Fig. 10a).

\nopagebreak
\begin{figure}[htb]
\begin{center}
\mbox{\epsfig{figure=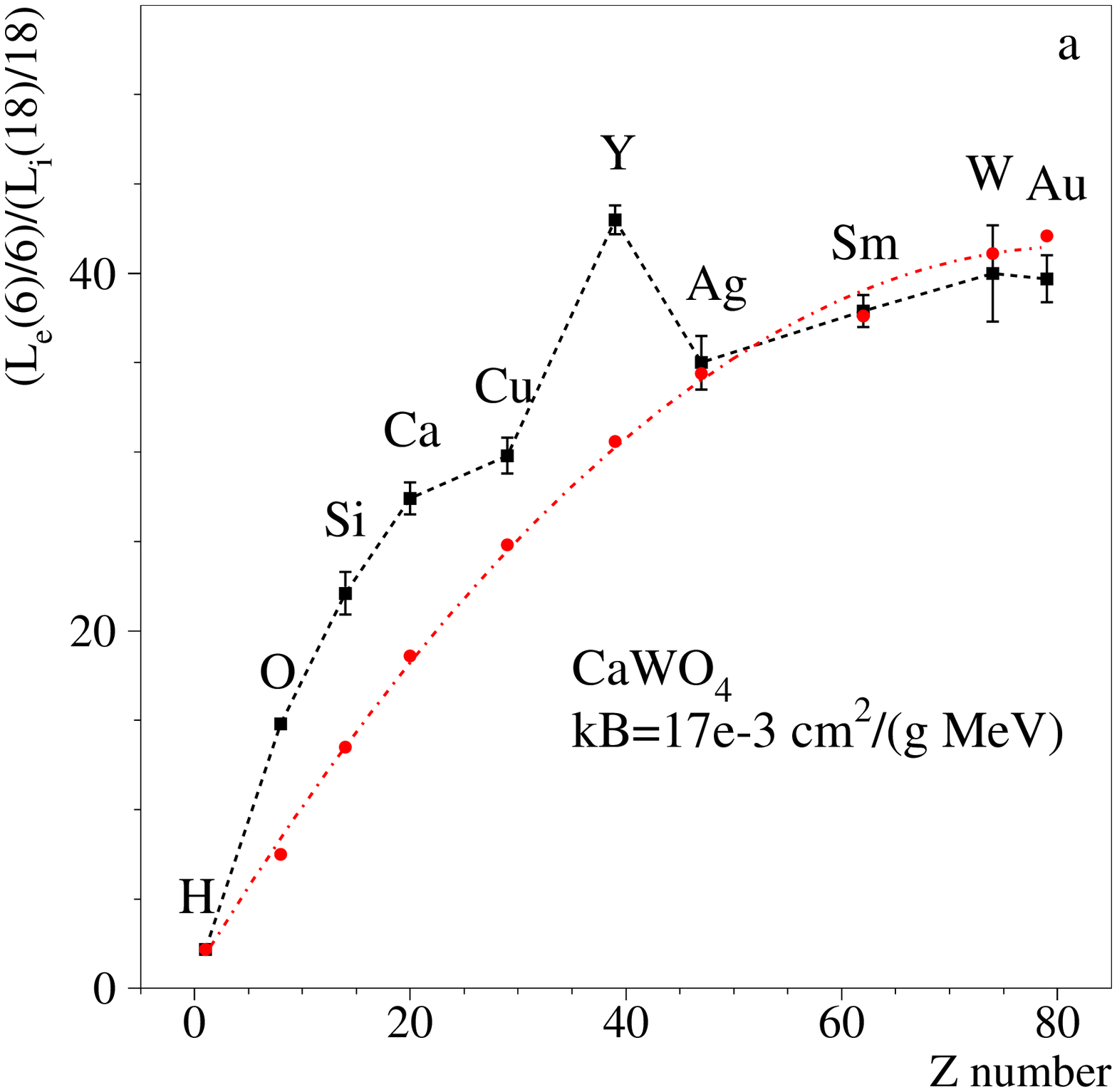,width=6.6cm}}~
\mbox{\epsfig{figure=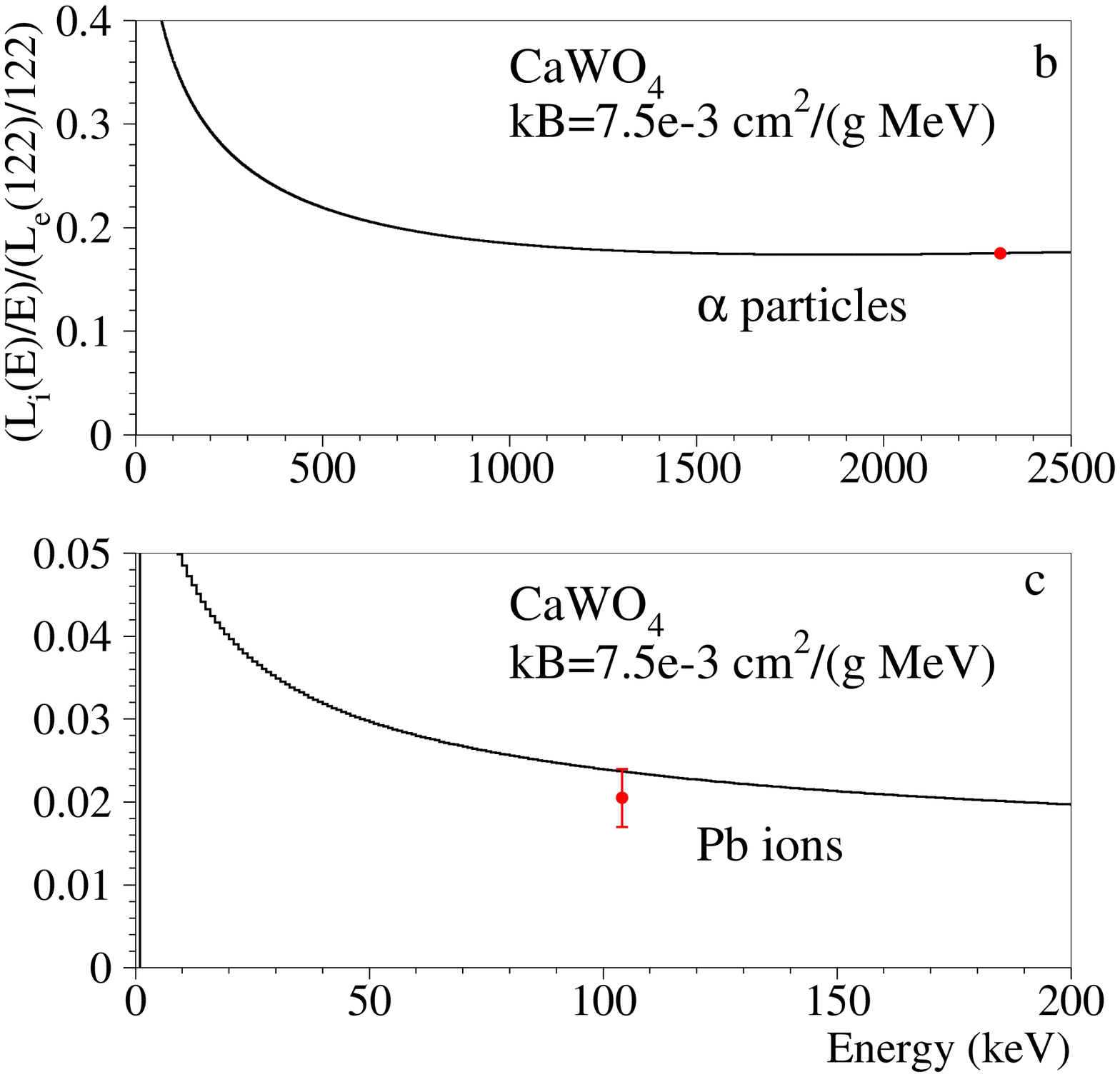,width=6.6cm}}
\caption{(Color online) (a) Dependence of inverse of the
relative light output $1/R_i$ at $E_i=18$ keV, normalized to that
for electron at $E_e=6$ keV on ion's $Z$ number: squares are
experimental points from \cite{Nin06}, and circles are calculated
values with $kB=17\times10^{-3}$ g MeV$^{-1}$ cm$^{-2}$. 
(b) Relative light output for $\alpha$ particles with
$kB=7.5\times10^{-3}$ g MeV$^{-1}$ cm$^{-2}$ to fit experimental
point at 2.3 MeV measured at 7 mK in \cite{Ang05}. (c) Predicted
curve for Pb ions with $kB$ the same as in (b) and experimental
point from \cite{Ang05} also measured at 7 mK.}
\end{center}
\end{figure}

Comparison of the calculated and experimental values gives extremely
high value of $\chi^2$/n.d.f. = 112 that is related mainly with deviations for
O, Si, Ca, Cu and especially Y points. It should be noted that the value for Y
evidently deviates from the general trend; also error bars in \cite{Nin06}
could be underestimated. F.e., new data for various ions are presented in Ref. 
\cite{Bav07} where points for O, Si, Ca, Cu and Y are much lower (and
in much better agreement with the calculated curve), while points for other
ions are approximately as in Fig. 10a. However, because the data of Ref.
\cite{Bav07} were preliminary and still are not explained in detail, we do
not use them here.

(5) Fig. 14 of Ref. \cite{Nin06} includes also $R'_i$ values for $\alpha$
particles at 2.3 MeV ($^{147}$Sm) and for Pb ions at 104 keV (nuclear recoil 
after $\alpha$ decay of $^{210}$Po) measured in \cite{Ang05}.
Taking into account that (a) quenching factors are energy dependent and (b) data for
$\alpha$ and Pb were measured at a temperature 7 mK \cite{Ang05},
one could not expect perfect agreement between these $R'_i$ and $R'_i$ for all other ions
taken in \cite{Nin06} at 18 keV and measured at a room temperature.
However, inside our ideology, data for $\alpha$ and Pb, taken at the same conditions,
should be self-consistent. Once more, we can use the relative light output for $\alpha$
particle $L_\alpha(E)/E$ at 2.3 MeV normalized to that for electrons at 122 keV
(as in Ref. \cite{Ang05}\footnote{However, numerically it is very close to quenching factor
$Q_i$, see Eqs. (5--7).}) to determine the Birks
factor;  it gives $kB=7.5\times10^{-3}$ g MeV$^{-1}$ cm$^{-2}$ (see Fig. 10b).
Curve with this $kB$ value for Pb ions is shown in Fig. 10c; at energy of 104 keV it
agrees with quenching factor given in \cite{Ang05,Nin06} ($\chi^2$/n.d.f. = 0.78).
Taking into account big difference in atomic numbers of $\alpha$ particle
and Pb ion (2 and 82, respectively), as well as in their energies (2.3 MeV and 104 keV),
this example demonstrates consistency in description of such diverse data.

\subsubsection{CsI(Tl) and CsI(Na)}

(1) Experimental data for quenching factors of Cs and I ions in
CsI(Tl) crystal scintillators ($\rho=4.51$ g cm$^{-3}$) published
in \cite{Pec99}, as well as in \cite{Par02} (KIMS
collaboration)\footnote{In Ref. \cite{Par02}, data with Tl
concentration of 0.128\% were chosen but values with other Tl
concentrations are similar.} and \cite{Wan02} (TEXONO
collaboration) are shown in Fig. 11a, 11b and 11c, respectively;
total energy range of ions was 7 to 135 keV. All these data sets
are well described by Eq. (5) with the same value of the Birks
factor: $kB=3.2\times10^{-3}$ g MeV$^{-1}$ cm$^{-2}$
($\chi^2$/n.d.f. value is 1.9, 0.51 and 0.49, respectively). 
Cs and I nuclei have very close atomic numbers and masses, and their
quenching factors also are very close (see Fig. 11a).

\nopagebreak
\begin{figure}[htb]
\begin{center}
\mbox{\epsfig{figure=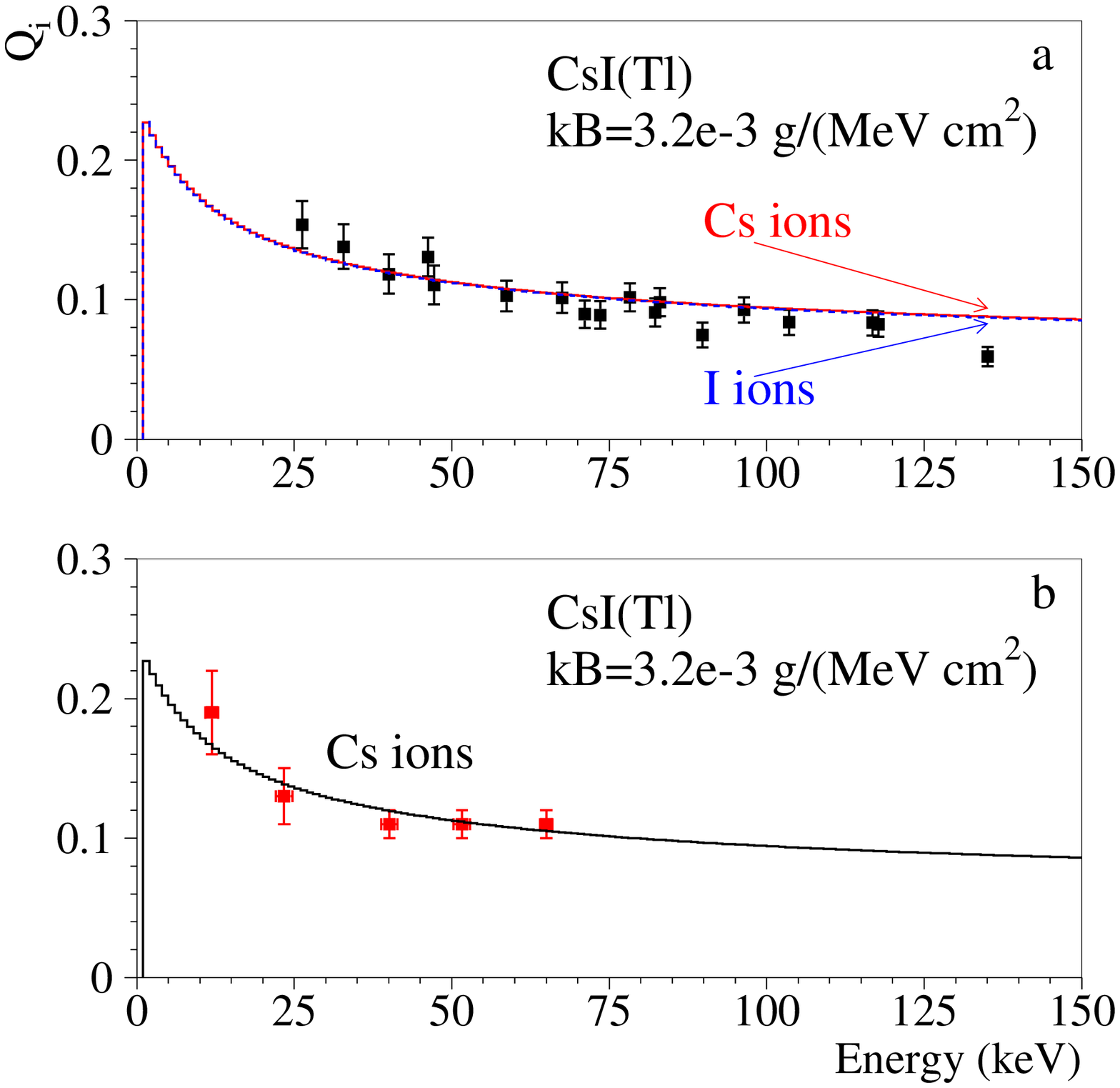,width=6.6cm}}~
\mbox{\epsfig{figure=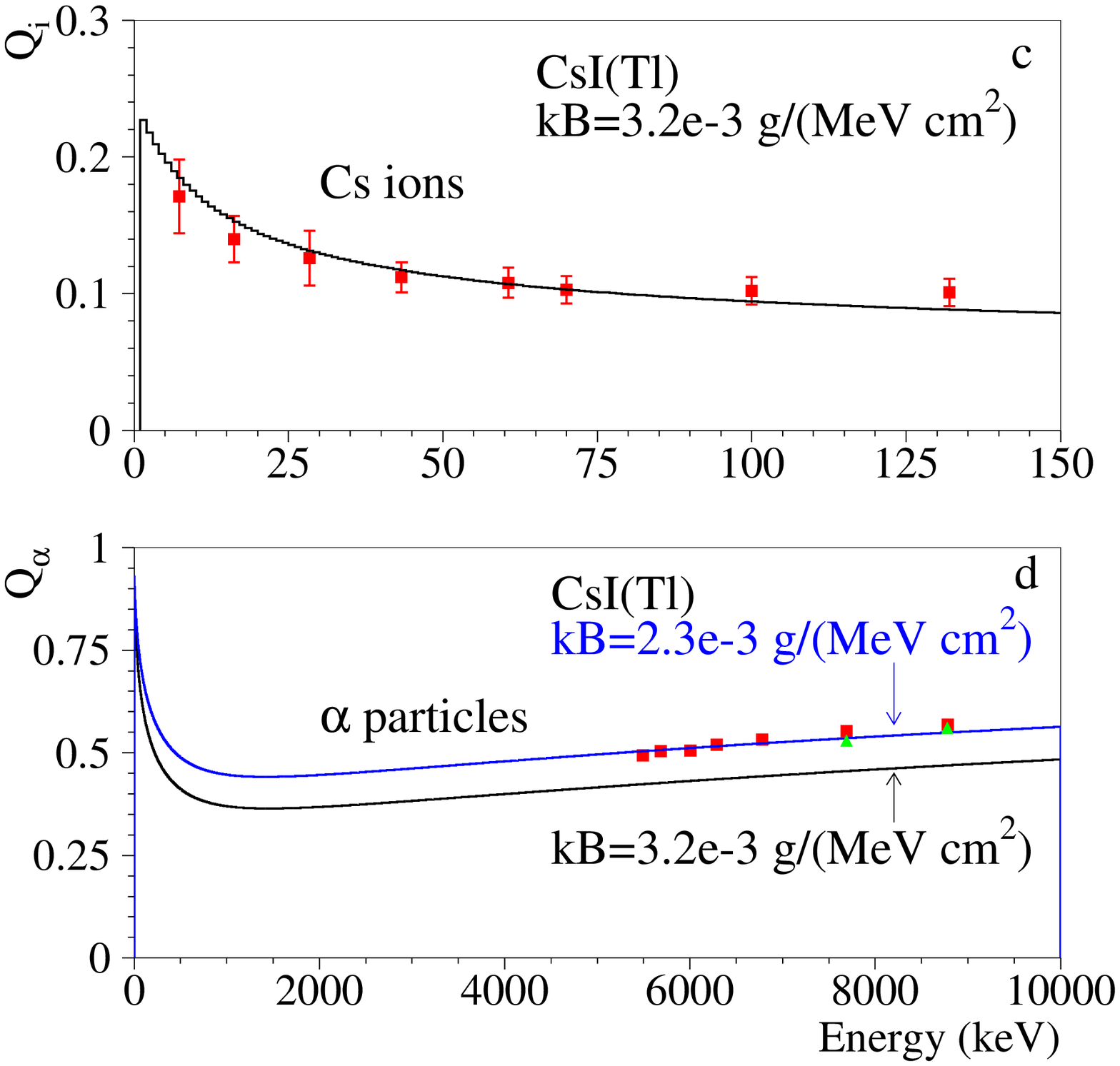,width=6.6cm}}
\caption{(Color online) 
Quenching factors (squares) for Cs and I ions in CsI(Tl) measured in:
\cite{Pec99} (a), \cite{Par02} (b), \cite{Wan02} (c) and their fit with
the Birks factor $kB=3.2\times10^{-3}$ g MeV$^{-1}$ cm$^{-2}$;
(d) quenching factors for $\alpha$ particles in CsI(Tl) measured in \cite{Kim03} (triangles)
and \cite{Zhu06} (squares) and calculated quenching
curves with $kB=3.2\times10^{-3}$ g MeV$^{-1}$ cm$^{-2}$ and
$kB=2.3\times10^{-3}$ g MeV$^{-1}$ cm$^{-2}$.}
\end{center}
\end{figure}

(2) The KIMS and TEXONO collaborations measured also quenching factors for $\alpha$
particles derived from studies of internal contamination of CsI(Tl) detectors
\cite{Kim03,Zhu06}. Their values are consistent; see Fig. 11d where
they are presented together with a calculated quenching curve for $\alpha$
particles with the same $kB=3.2\times10^{-3}$ g MeV$^{-1}$ cm$^{-2}$ as
above for Cs and I ions. However, this curve is lower than the experimental points
which are much better described by curve with
$kB=2.3\times10^{-3}$ g MeV$^{-1}$ cm$^{-2}$.
This could be some underestimation of $Q_\alpha$ values in our approach but as well
we have to remember that the data for Cs and I were taken in devoted
measurements with neutron beams while data for $\alpha$ particles were collected
in separate measurements with probably non-identical experimental conditions and
details of data treatment (temperature in the Cs/I measurements by KIMS was
24.5$^\circ$C in \cite{Par02}, 
and $26-29$$^\circ$C in $\alpha$ measurements \cite{Kim03}, etc.).

(3) Experimental quenching factors for Cs and I ions in CsI(Na) crystal scintillators 
measured in \cite{Par02} are shown in Fig. 12a (data for 0.0188\% of Na 
were taken but results for other Na amounts are similar).  
Data points are fitted by Eq. (5) calculated with 
$kB=5.5\times10^{-3}$ g MeV$^{-1}$ cm$^{-2}$ ($\chi^2$/n.d.f. = 1.9).
Quenching factors for $\alpha$ particles obtained with this $kB$ value are shown
in Fig. 12b.

\nopagebreak
\begin{figure}[htb]
\begin{center}
\mbox{\epsfig{figure=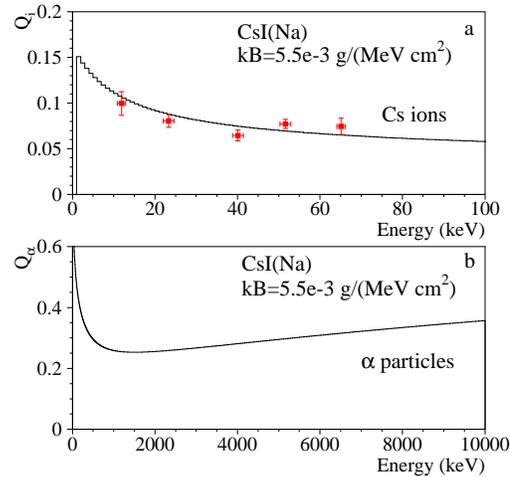,width=6.6cm}}
\caption{(Color online) 
(a) Quenching factors for Cs and I ions in CsI(Na) measured in
\cite{Par02} and their fit by Eq. (5) with
$kB=5.5\times10^{-3}$ g MeV$^{-1}$ cm$^{-2}$.
(b) Quenching factors for $\alpha$ particles in CsI(Na)
calculated with the $kB$ value as in (a).}
\end{center}
\end{figure}

\subsubsection{NaI(Tl)}

(1) Quenching factors for Na and I nuclei in NaI(Tl) crystal scintillators
($\rho=3.67$ g cm$^{-3}$) measured in Ref. \cite{Tov98} in energy range
of $7-215$ keV and $13-54$ keV, respectively, are presented in Fig. 13a.
Value of the Birks factor which allows to describe curve for Na ions with Eq. (5)
is equal: $kB=3.8\times10^{-3}$ g MeV$^{-1}$ cm$^{-2}$ ($\chi^2$/n.d.f. = 1.2).
Once found, it also allows to calculate quenching curve for I ions, and this curve
is in excellent agreement ($\chi^2$/n.d.f. = 0.26)
with the experimental data, as one can see in Fig. 13a.

\nopagebreak
\begin{figure}[htb]
\begin{center}
\mbox{\epsfig{figure=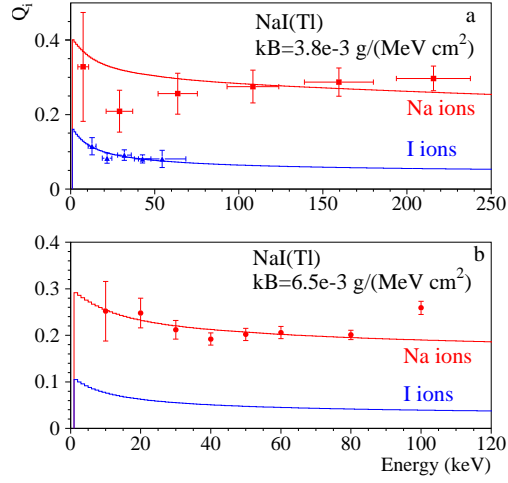,width=6.6cm}}
\caption{(Color online) 
(a) Quenching factors for Na (squares) and I (triangles) in NaI(Tl)
measured in \cite{Tov98} and calculated curves with
$kB=3.8\times10^{-3}$ g MeV$^{-1}$ cm$^{-2}$.
(b) $Q_i$ values for Na ions from \cite{Cha08} and fitting curve with
$kB=6.5\times10^{-3}$ g MeV$^{-1}$ cm$^{-2}$. Prediction for I ions is also shown.}
\end{center}
\end{figure}

(2) In recent work \cite{Cha08}, quenching factors for Na ions in NaI(Tl) were
measured in more detail in low energy region ($10-100$ keV). Obtained experimental
values are lower than those in \cite{Tov98}. This demands higher $kB$ factor
to describe higher quenching: $kB=6.5\times10^{-3}$ g MeV$^{-1}$ cm$^{-2}$.
Corresponding calculated curve well describes experimental values (except of
the last point at 100 keV), see Fig. 13b ($\chi^2$/n.d.f. is equal 4.0 for
all the points and 0.56 with the last point excluded). 
Calculations for I ions are also presented
(however quenching factors for I were not measured in \cite{Cha08}).

It is worth noting that with the $kB$ values as in Figs. 13a and 13b, our suggested 
procedure predicts a quenching factor for 6 MeV $\alpha$ particles of 0.35 and 0.25, 
respectively; these values are lower by a factor 1.5 -- 2 than those typically measured
(however, under different experimental conditions) for NaI(Tl) (see f.e. \cite{Mat56}).

(3) Quenching factors for Na and I ions in NaI(Tl) scintillators were also measured
in other works with monoenergetic neutron beams as: \\
-- $Q_{\mathrm{Na}}(E\simeq18-800$ keV$)\simeq0.30$, and
   $Q_{\mathrm{I}}(E\simeq20-120$ keV$)\simeq0.1$
(with some increase at lower energies) \cite{Spo94}; \\
-- $Q_{\mathrm{Na}}(E\simeq18-74$ keV$)=0.25\pm0.03$, and
   $Q_{\mathrm{I}}(E\simeq40-100$ keV$)=0.08\pm0.02$ \cite{Ger99}; \\
-- $Q_{\mathrm{Na}}(E\simeq50-336$ keV), values are consistent with constant of
   $0.27\pm0.02$ \cite{Sim03}.

In addition, in measurements with $^{252}$Cf source quenching factors
in some effective energy range were obtained as: \\
-- $Q_{\mathrm{Na}}(E=5-100$ keV$)=0.40\pm0.20$, and
   $Q_{\mathrm{I}}(E\simeq40-300$ keV$)=0.05\pm0.02$ \cite{Fus93}; \\
-- $Q_{\mathrm{Na}}(E=7-100$ keV$)=0.30$, and
   $Q_{\mathrm{I}}(E\simeq20-330$ keV$)=0.09$ \cite{Ber96}.

Comparing energy ranges investigated in these works with behaviour of quenching
curves in Figs. 13a,b, one can note that the energy thresholds were not low enough to
observe increase of quenching factors at lower energies
which is predicted in our approach, and at higher energies
$Q_i$ are consistent with constant values (taking into account experimental
uncertainties). However, works \cite{Cha08,Tov98} with lower thresholds
give experimental values consistent with this prediction.

(4) In Ref. \cite{Ber08b}, energy dependence of quenching factor for $\alpha$ particles
in range of energies $E_\alpha=5.7-6.8$ MeV
(from internal contamination of one of NaI(Tl) crystals) was obtained as:
$Q_\alpha(E_\alpha) = 0.467(6)+0.0257(10)\times E_\alpha$, where $E_\alpha$ is in MeV.
This energy dependence (shown in Fig. 14a) can be reproduced by Eq. (5) with the
Birks factor $kB=1.25\times10^{-3}$ g MeV$^{-1}$ cm$^{-2}$, much lower than those
found for data of Refs. \cite{Cha08,Tov98} in Figs. 13a,b.

\nopagebreak
\begin{figure}[htb]
\begin{center}
\mbox{\epsfig{figure=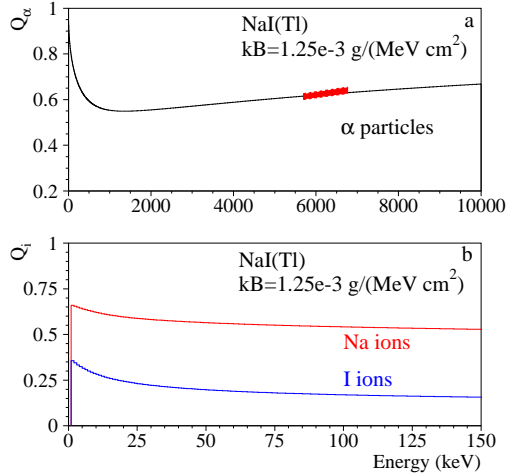,width=6.6cm}}
\caption{(Color online) 
(a) Quenching factor for $\alpha$ particles measured in \cite{Ber08b}
(filled area represents corridor of uncertainties) and its fit with
$kB=1.25\times10^{-3}$ g MeV$^{-1}$ cm$^{-2}$.
(b) Predicted quenching curves for Na and I ions with
$kB=1.25\times10^{-3}$ g MeV$^{-1}$ cm$^{-2}$.
Note that the calculated $Q_{\mathrm{Na}}$ and $Q_{\mathrm{I}}$ are $\simeq2$ higher
than those measured in the DAMA \cite{Ber96} and other experiments -- however, under
different conditions.}
\end{center}
\end{figure}

Following our method, we predict quenching factors for Na and I ions in this crystal
with the obtained $kB$ value; they are shown in Fig. 14b and are significantly higher 
(around 0.6 for Na, and 0.3 for I) than the ones measured by the DAMA group in \cite{Ber96}
(0.30 and 0.09, respectively), which are also similar to other determinations 
available in literature. 

Considering that our suggested procedure predicts: 
i) for the cases of Refs. \cite{Cha08,Tov98} a quenching factor for $\alpha$ 
particles in the MeV region lower by a factor 1.5 -- 2 than those typically measured; 
ii) quenching factors for Na and I ions at low energy higher by a factor about 
2 than those measured by \cite{Ber96} and other experiments; 
one could consider the obtained here results with some cautious attitude.
However, we want to once more remind that quenching factors for Na/I ions and
$\alpha$ particles were measured in both cases under different experimental 
conditions. It would be of great interest to measure them simultaneously, with
the same data taking and treatment. 
Fig. 10 for CaWO$_4$ and Fig. 13a for NaI(Tl) give examples that, if such a 
condition is fulfilled, the method gives self-consistent description of data for 
ions with very different ($A,Z$) values.

As general consideration, it is worth noting that higher quenching 
factors would always be of big importance for searches of dark matter 
particles because, with a fixed energy threshold of a detector, higher 
quenching factors allow to study lower energies of DM particles. 
For example, when assuming some cases for WIMPs, the energy distribution 
would drop quasi-exponentially with energy and, thus, this would lead to 
higher experimental sensitivities in the DM searches.

\subsubsection{CeF$_3$}

Quenching factors for $\alpha$ particles in CeF$_3$ crystal scintillator
($\rho=6.16$ g cm$^{-3}$) were studied in \cite{Bel03} in the range of energies
$E_\alpha=2.1-8.8$ MeV. They are shown in Fig. 15 together with fit by Eq. (5) with
$kB=11.1\times10^{-3}$ g MeV$^{-1}$ cm$^{-2}$ ($\chi^2$/n.d.f. = 1.8).

\nopagebreak
\begin{figure}[htb]
\begin{center}
\mbox{\epsfig{figure=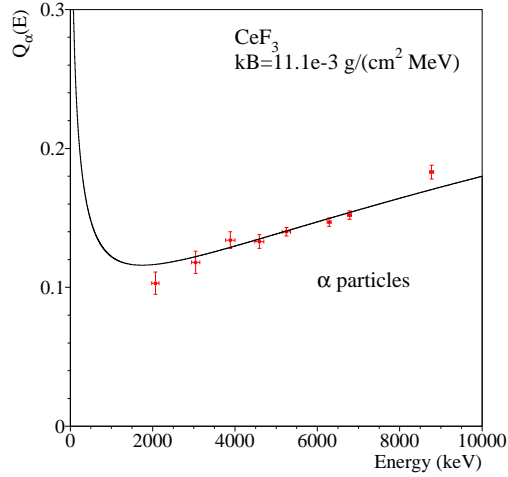,width=6.6cm}}
\caption{(Color online) 
Quenching factors for $\alpha$ particles in CeF$_3$ measured
in \cite{Bel03} and their fit with $kB=11.1\times10^{-3}$ g MeV$^{-1}$ cm$^{-2}$.}
\end{center}
\end{figure}

\subsection{Liquid noble gases}

\subsubsection{Liquid Xe}

(1) Some data on quenching factors in liquid Xe scintillator (LXe, $\rho=3.52$ g cm$^{-3}$
at $-109^\circ$ C \cite{Lid03}\footnote{Density of liquid Xe depends on pressure and 
temperature. However, numerical value of the Birks factor $kB$ 
does not depend on value of density if $kB$ is measured in g MeV$^{-1}$ cm$^{-2}$.})
also support predicted in the described approach increase of $Q_i$ values at low energies.
Results obtained in Ref. \cite{Ber98} are shown in Fig. 16a; they are well described
by Eq. (5) with the Birks factor $kB=3.5\times10^{-4}$ g MeV$^{-1}$ cm$^{-2}$
($\chi^2$/n.d.f. = 0.36).
On the contrary, the Lindhard's theory \cite{Lin63}
predicts decrease of $Q_i$ values at low energies;
it is also shown in Fig. 16a (calculated using description in \cite{Mei08}).

\nopagebreak
\begin{figure}[htb]
\begin{center}
\mbox{\epsfig{figure=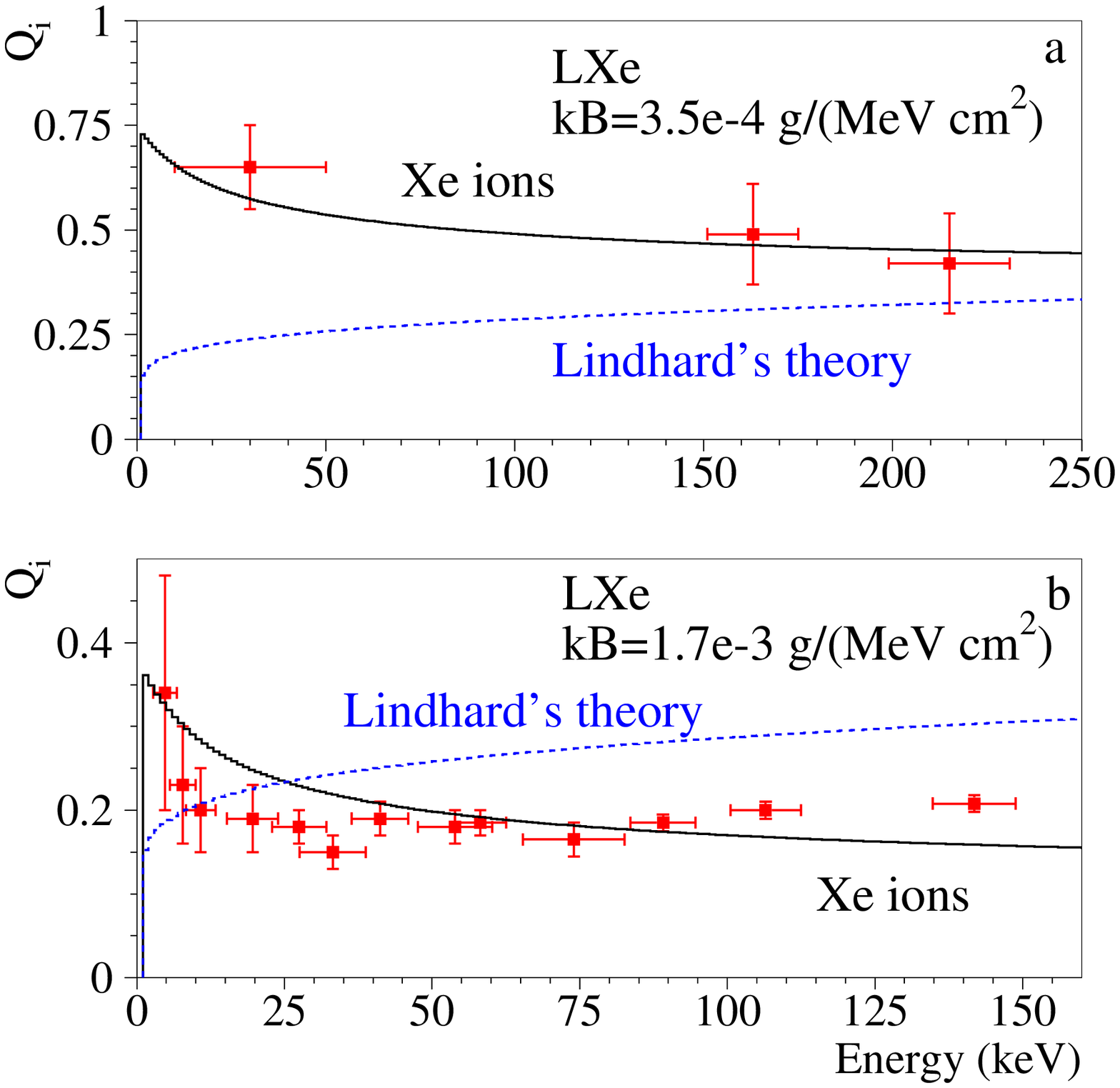,width=6.6cm}}~
\mbox{\epsfig{figure=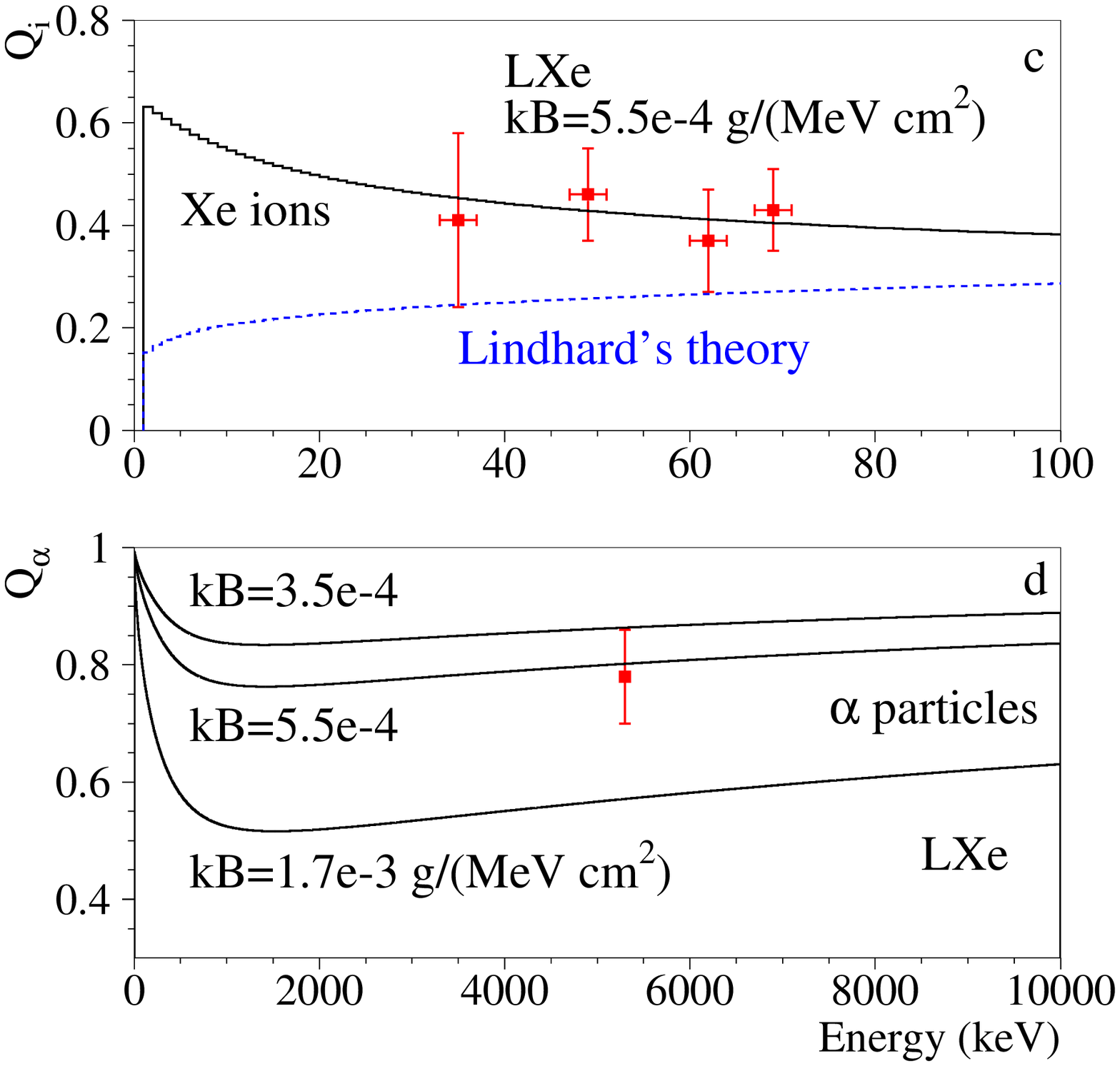,width=6.6cm}}
\caption{(Color online) Quenching factors for Xe ions in LXe:
(a) measured in \cite{Ber98} and fitting curve with 
$kB=3.5\times10^{-4}$ g MeV$^{-1}$ cm$^{-2}$;
(b) data from \cite{Che06} and fitting curve with
$kB=1.7\times10^{-3}$ g MeV$^{-1}$ cm$^{-2}$;
(c) measured in \cite{Ber01} and calculated curve with
$kB=5.5\times10^{-4}$ g MeV$^{-1}$ cm$^{-2}$.
Lindhard's theoretical prediction is shown by dashed line in (a)-(c).
(d) $Q_\alpha$ quenching curves calculated with different $kB$ values
together with experimental point for 5.3 MeV $\alpha$ particle from \cite{Tan01}.}
\end{center}
\end{figure}

(2) Quenching factors for Xe ions in LXe measured in \cite{Che06}\footnote{More exactly,
relative light outputs $L_i(E)/E$ normalized to that for electrons of 122 keV were
measured in \cite{Che06}; however they are very close here to quenching factors, see
Eqs. (5--7).} are shown in Fig. 16b.
They are much lower than those of Ref. \cite{Ber98}, and to reproduce these
data with high quenching, the Birks factor also should have bigger value:
$kB=1.7\times10^{-3}$ g MeV$^{-1}$ cm$^{-2}$
($\chi^2$/n.d.f. = 5.0).

(3) Experimental data from \cite{Ber01}, measured in not so wide energy range
($35 - 69$ keV), are consistent with a constant value (Fig. 16c). 
However, as well they
are in good agreement with description by Eq. (5) with 
the Birks factor $kB=5.5\times10^{-4}$ g MeV$^{-1}$ cm$^{-2}$
($\chi^2$/n.d.f. = 0.15).

(4) Quenching curves for $\alpha$ particles calculated with the three above quoted 
$kB$ values are shown in Fig. 16d together with experimental result of Ref. 
\cite{Tan01} for 5.3 MeV $\alpha$ particles: $Q_\alpha=0.78\pm0.08$.
Because data of Ref. \cite{Tan01} were obtained in experimental
conditions different from those in \cite{Ber98,Che06,Ber01}, no surprise that 
not all curves are in good agreement with the experimental point:
to be in agreement, quenching factors for all particles
should be obtained with the same detector and in the same experimental conditions and
data treatment.

(5) It should be noted also that there are other experimental data sets for quenching
factors of Xe ions in LXe which, on contrary, demonstrate decrease of $Q_i$ values
at lower energies \cite{Apr09}. 
Results of Refs. \cite{Arn00,Aki02} were measured in energy range of $\simeq 47-110$
keV and $\simeq 43-65$ keV, respectively, and are not far from constant values, as
also could be expected from Figs. 16a-c.

\subsubsection{Liquid Ar}

Data for liquid argon ($\rho=1.40$ g cm$^{-3}$) are rather scarce.
In fact, only one measurement was performed in the framework of
the WARP project \cite{Bru05} where quenching factor for Ar ions
was obtained with monoenergetic neutrons of 14 MeV, however
without fixing the angle of neutron scattering. Thus quenching
factor was obtained for some effective energy range, with mean
value of 65 keV, as: $Q_{\mathrm{Ar}}(E\simeq65$
keV$)=0.28\pm0.03$ (at applied electric field of 1 kV/cm). This
point can be reproduced by Eq. (5) with the Birks factor of
$kB=1.25\times10^{-3}$ g MeV$^{-1}$ cm$^{-2}$; see Fig. 17a, where
also curve calculated in the Lindhard's theory \cite{Lin63,Mei08}
is shown.

\nopagebreak
\begin{figure}[htb]
\begin{center}
\mbox{\epsfig{figure=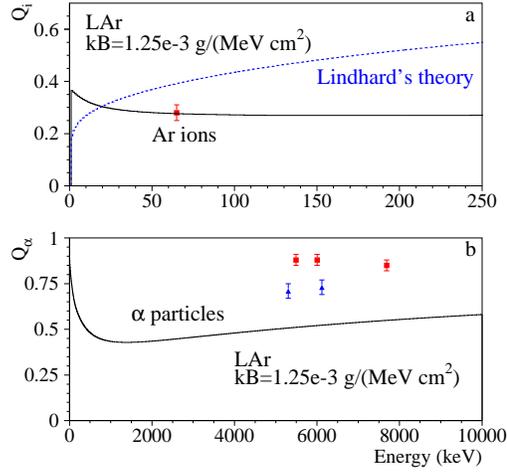,width=6.6cm}}
\caption{(Color online) 
(a) Quenching factor for Ar ions in LAr measured in \cite{Bru05}
and fitting curve with $kB=1.25\times10^{-3}$ g MeV$^{-1}$ cm$^{-2}$.
Lindhard's theoretical prediction is shown by dashed line.
(b) $Q_\alpha$ quenching curve for LAr calculated with
$kB=1.25\times10^{-3}$ g MeV$^{-1}$ cm$^{-2}$ and experimental data
measured in \cite{Hit87} (triangles) and \cite{Pei08} (squares).}
\end{center}
\end{figure}

Dependence of quenching factor for $\alpha$ particles on energy, calculated with this
$kB$ value, is shown in Fig. 17b. Experimental values of $Q_\alpha$ measured in
experiments \cite{Hit87,Pei08} also are given; however, because they were obtained
in different experimental conditions, they are not obliged to lay on the calculated
curve with $kB=1.25\times10^{-3}$ g MeV$^{-1}$ cm$^{-2}$, and even are not obliged
to be in agreement between themselves (as one can see in Fig. 17b).

\subsubsection{Liquid Ne}

Experimental situation for liquid neon ($\rho=1.21$ g cm$^{-3}$) 
is even worse than for LAr:
to-date is only one experiment \cite{Nik08} where the relative light output
normalized to that for electrons of $E_0=511$ keV was measured at the energy $E=387\pm11$
keV as: $(L_{\mathrm{Ne}}(E)/E)/(L_e(E_0)/E_0)=0.26\pm0.03$.
Calculated quenching curve of Eq. (6) which is normalized to this experimental point
is shown in Fig. 18a ($kB=2.0\times10^{-3}$ g MeV$^{-1}$ cm$^{-2}$); Fig. 18b
shows prediction for quenching for $\alpha$ particles calculated with this $kB$ factor.

\nopagebreak
\begin{figure}[htb]
\begin{center}
\mbox{\epsfig{figure=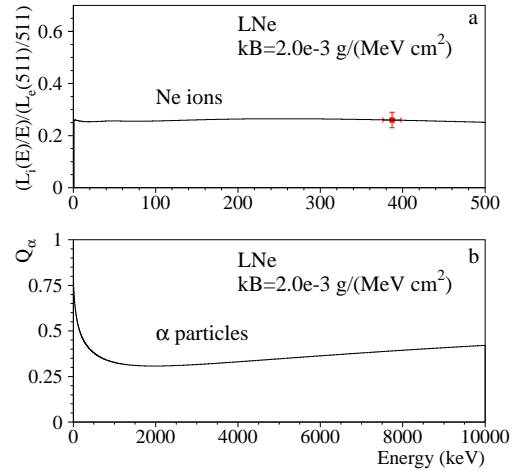,width=6.6cm}}
\caption{(Color online) (a) Relative light output for Ne ions in liquid Ne. 
Curve is normalized to experimental
value at 387 keV from Ref. \cite{Nik08}
($kB=2.0\times10^{-3}$ g MeV$^{-1}$ cm$^{-2}$).
(b) Predicted quenching curve for $\alpha$'s in LNe with
$kB=2.0\times10^{-3}$ g MeV$^{-1}$ cm$^{-2}$.}
\end{center}
\end{figure}

\subsection{Estimation of quenching factors for nuclear recoils in some scintillators}

While quenching factors for nuclear recoils that constitute such scintillators
as C$_9$H$_{12}$, CaF$_2$(Eu), ZnWO$_4$, CaWO$_4$, CsI(Tl), NaI(Tl), LXe, LAr, LNe
can be found above, for some other scintillators, which are considered as perspective 
detectors in the DM searches, quenching factors were not measured to-date. 
Below we give estimation of $Q_i$ values for nuclear recoils in CdWO$_4$, PbWO$_4$, CeF$_3$, 
Bi$_4$Ge$_3$O$_{12}$, LiF and ZnSe scintillators which are based on measured
quenching factors for $\alpha$ particles in these materials. 
We should remember, of course, that the $kB$ values and thus quenching factors
for $\alpha$ particles and recoils can be different for the same material
in different conditions of measurements and data treatment. 
Nevertheless, results given below could be useful as providing some initial values
of $Q_i$.

(1) Quenching factors for O, Cd and W ions in CdWO$_4$ scintillator are shown in 
Fig. 19a for two extreme $kB$ values: 
$kB=10.1\times10^{-3}$ g MeV$^{-1}$ cm$^{-2}$ and
$kB=21.5\times10^{-3}$ g MeV$^{-1}$ cm$^{-2}$. 
The first $kB$ value was obtained by fitting experimental data of \cite{Dan03} 
for $\alpha$ particles (see Fig. 4a), and the second one by fitting data 
of \cite{Faz98} for $\alpha$ particles and protons (Fig. 4b).

\nopagebreak
\begin{figure}[htbp]
\begin{center}
\mbox{\epsfig{figure=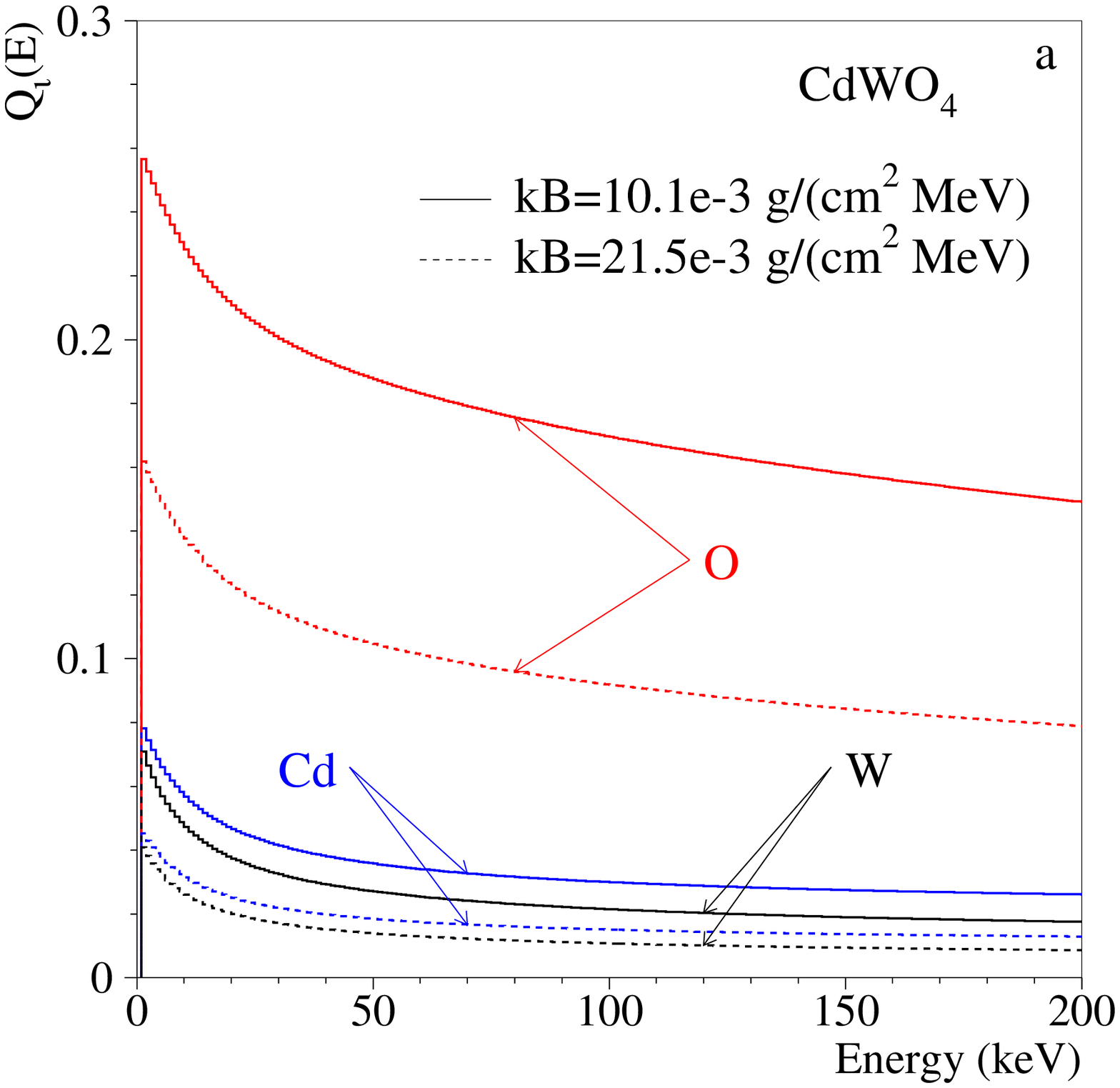,width=6.6cm}}~
\mbox{\epsfig{figure=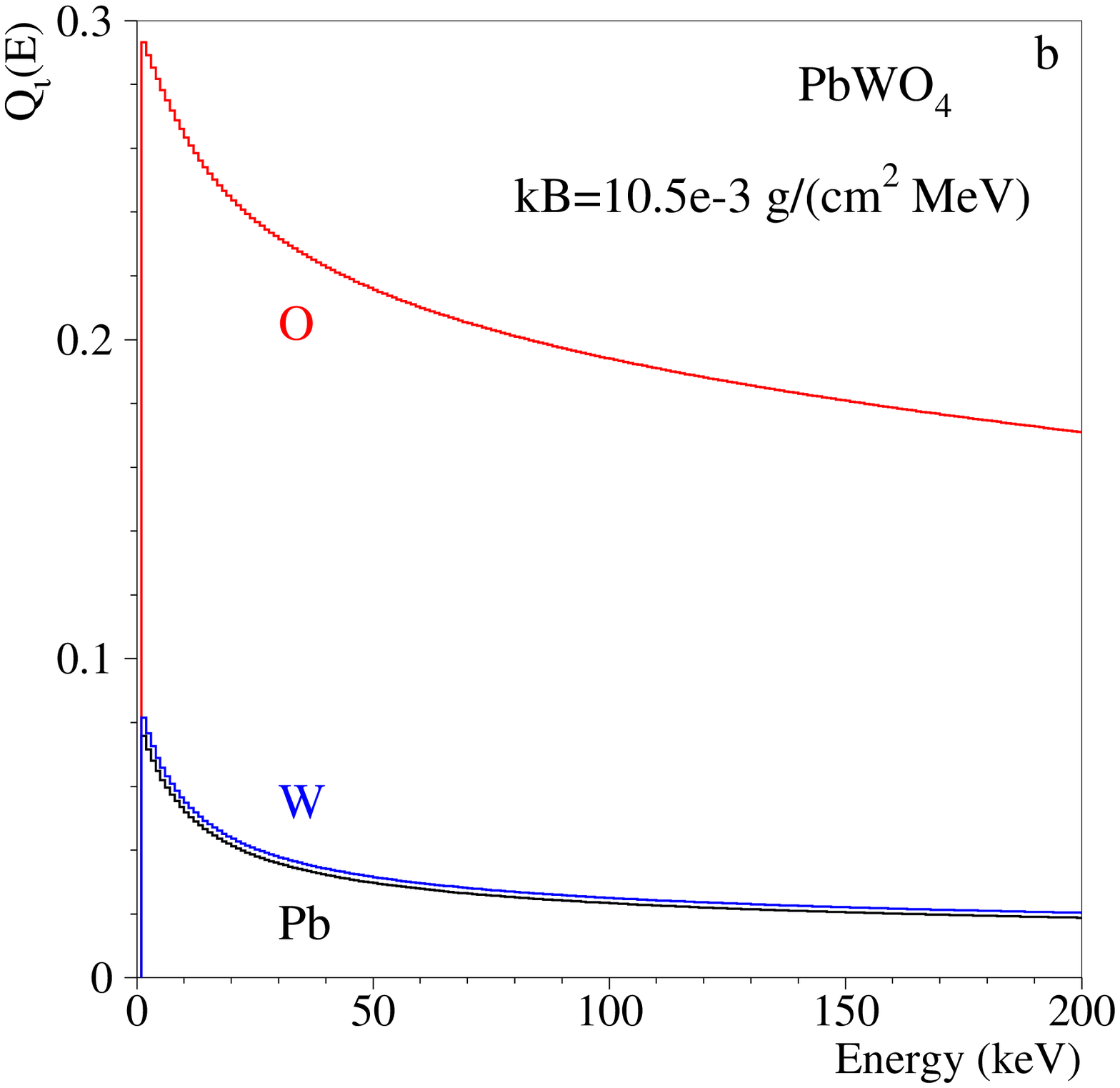,width=6.6cm}} \\
\mbox{\epsfig{figure=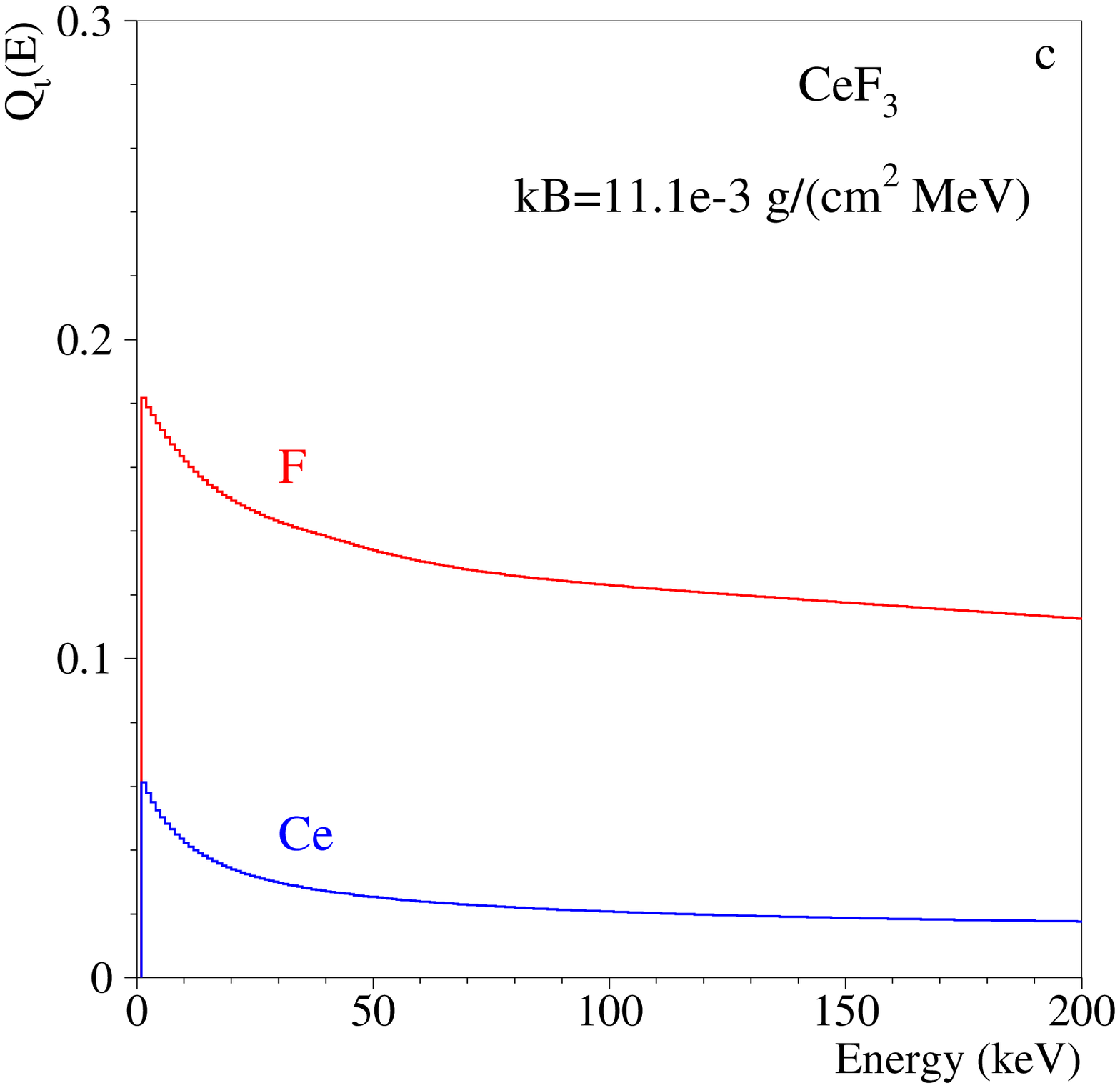,width=6.6cm}}~
\mbox{\epsfig{figure=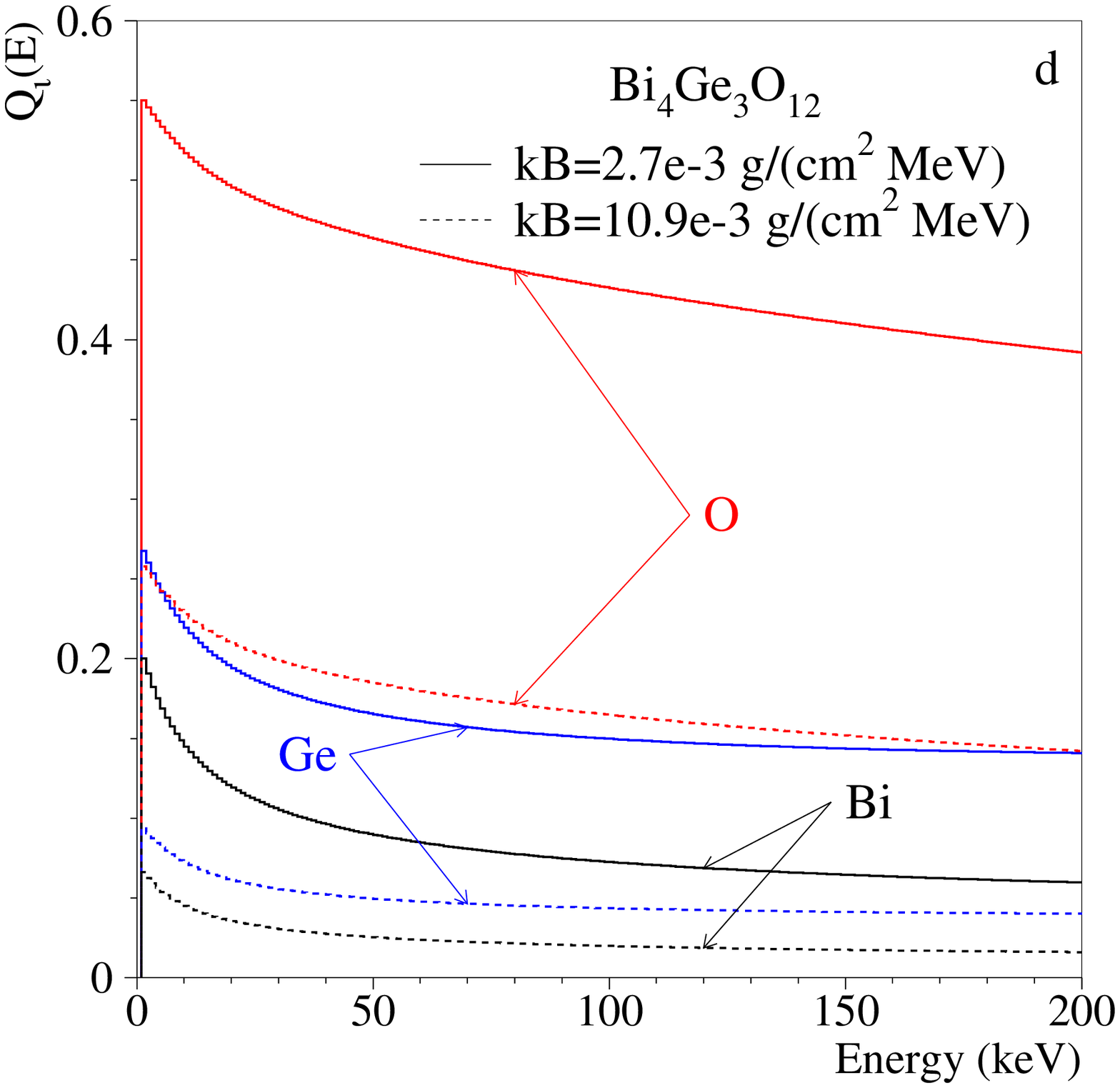,width=6.6cm}} \\
\mbox{\epsfig{figure=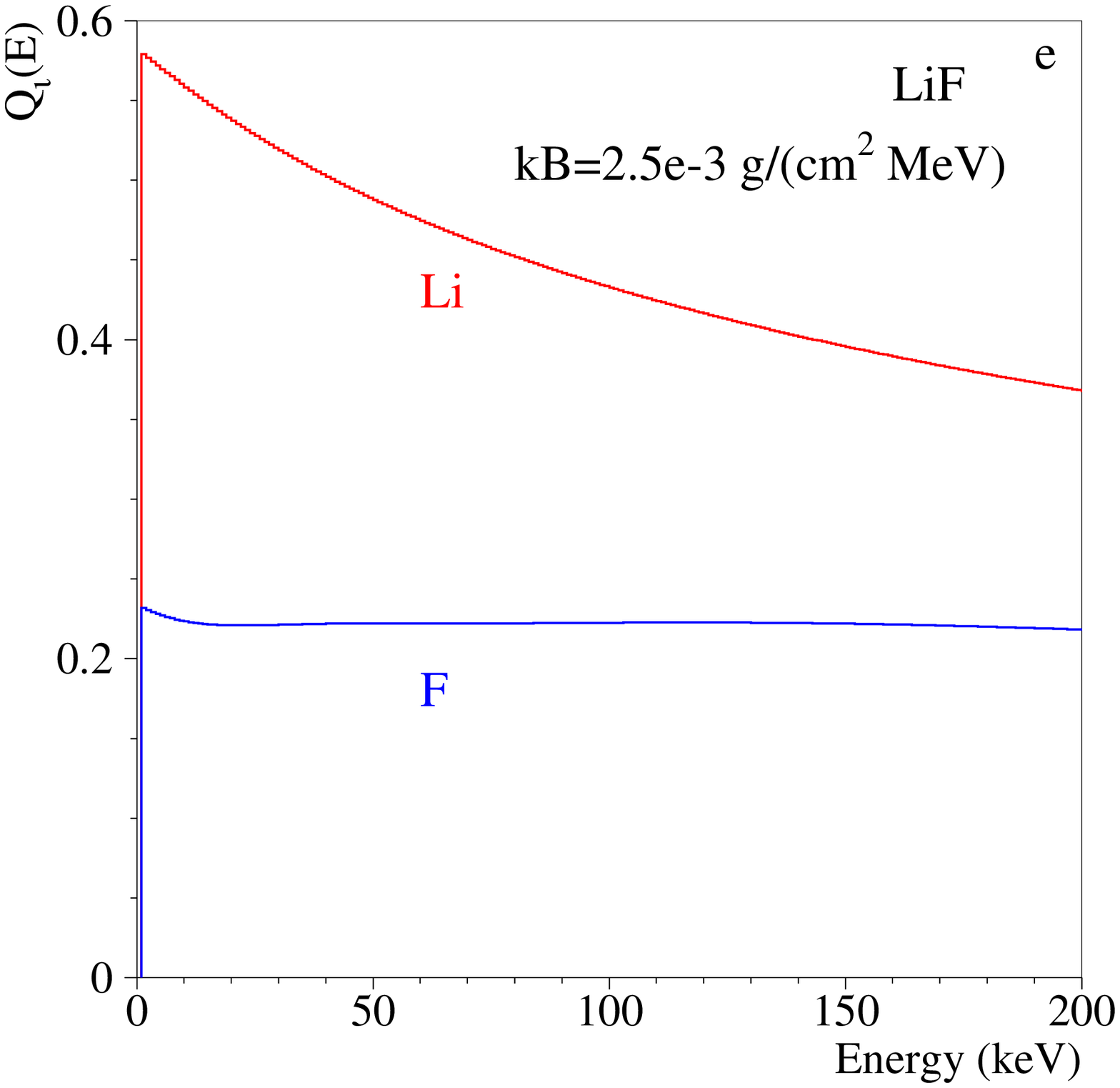,width=6.6cm}}~
\mbox{\epsfig{figure=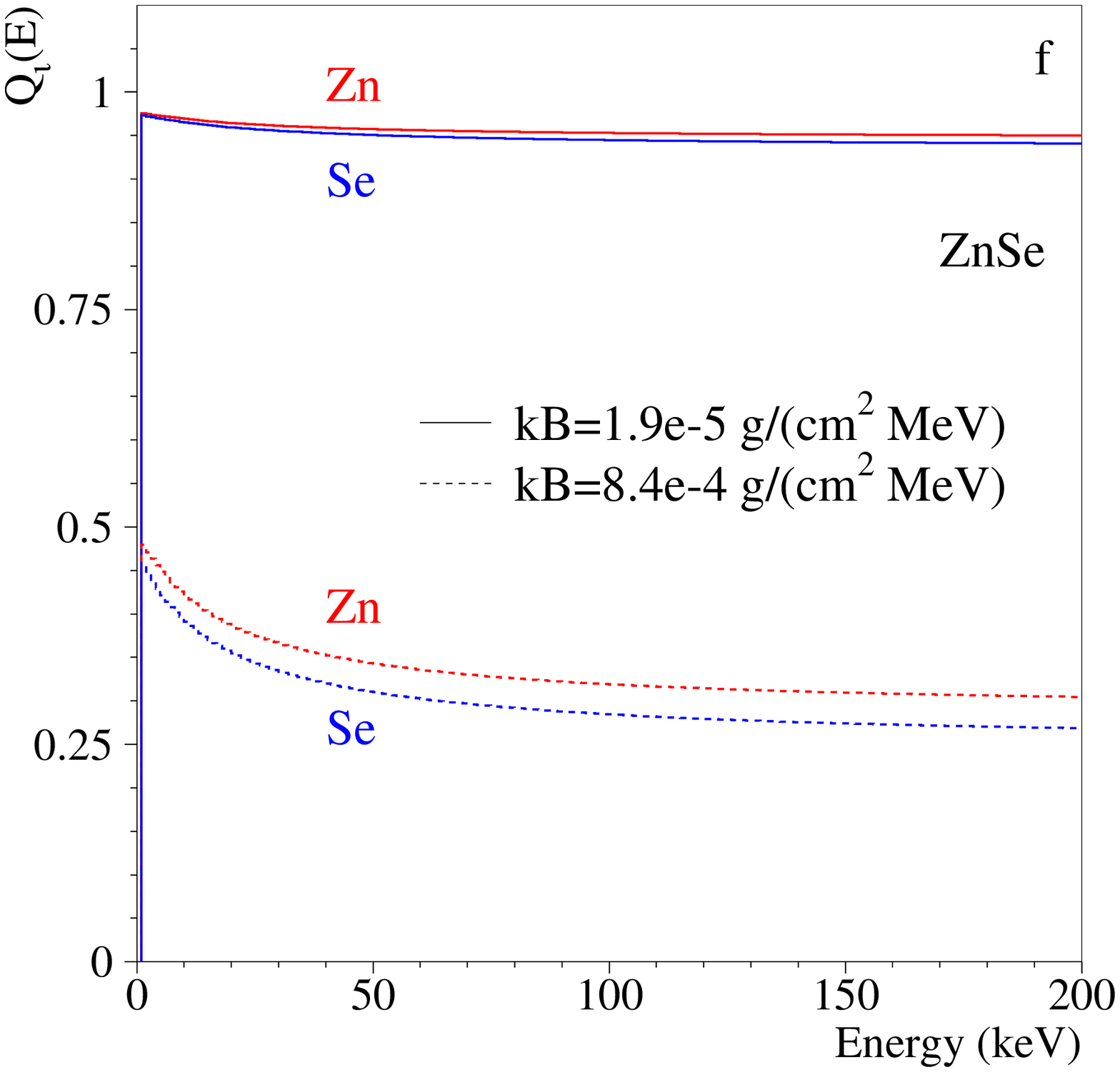,width=6.6cm}}
\caption{(Color online) 
Quenching factors for nuclear recoils in scintillators:
(a) O, Cd and W ions in CdWO$_4$;
(b) O, W and Pb ions in PbWO$_4$;
(c) F and Ce ions in CeF$_3$;
(d) O, Ge and Bi ions in Bi$_4$Ge$_3$O$_{12}$;
(e) Li and F ions in LiF;
(f) Zn and Se ions in ZnSe.
}
\end{center}
\end{figure}

(2) To calculate $Q_i$ for O, W and Pb ions in PbWO$_4$ (Fig. 19b), 
we use the value of
$kB=10.5\times10^{-3}$ g MeV$^{-1}$ cm$^{-2}$ obtained by fitting data
of \cite{Dan06} for $\alpha$ particles (see Fig. 6).

(3) Quenching factors for F and Ce nuclear recoils in CeF$_3$ scintillator 
(Fig. 19c) are
calculated supposing $kB=11.1\times10^{-3}$ g MeV$^{-1}$ cm$^{-2}$
obtained by fitting data of \cite{Bel03} for $\alpha$ particles (see Fig. 15).

(4) For Bi$_4$Ge$_3$O$_{12}$ crystal scintillator ($\rho=7.13$ g cm$^{-3}$), 
not only quenching factors for
nuclear recoils were not measured, but also quenching for $\alpha$ particles
was not studied in detail. $Q_\alpha$ values were quoted in several works, 
but mainly for $\simeq5.5$ MeV $\alpha$ particles from $^{241}$Am source. 
The results for $E_\alpha=5.5$ MeV are quite different:
$Q_\alpha=0.17$ at 20 mK temperature \cite{Cor04},
$Q_\alpha=0.20-0.21$ at a room temperature \cite{Sys98}, and
$Q_\alpha=0.45$ at 12 mK temperature \cite{Meu99}.
Result of $Q_\alpha=0.30\pm0.03$ for $E_\alpha=5.5-8.8$ MeV \cite{Dlo92}
is also known.

We give in Fig. 19d calculations of quenching factors for O, Ge and Bi recoils in 
Bi$_4$Ge$_3$O$_{12}$ scintillator for two extreme $kB$ values:
$kB=2.7\times10^{-3}$ g MeV$^{-1}$ cm$^{-2}$ (which reproduces value of  
$Q_\alpha=0.45$ for 5.5 MeV $\alpha$ particle of \cite{Meu99}) and
$kB=10.9\times10^{-3}$ g MeV$^{-1}$ cm$^{-2}$ (which reproduces   
$Q_\alpha=0.17$ for 5.5 MeV $\alpha$ particle \cite{Cor04}).

(5) Information on quenching in LiF crystal scintillator ($\rho=2.64$ g cm$^{-3}$)
is extremely scarce. We were able to find only one paper where value of
$Q_\alpha=0.29$ was measured for 5.5 MeV $\alpha$ particles \cite{Cor04}.
This value can be reproduced by Eq.~(5) with the Birks factor
$kB=2.5\times10^{-3}$ g MeV$^{-1}$ cm$^{-2}$. 
Quenching factors for Li and F ions in LiF with this $kB$ are shown in Fig. 19e.

(6) ZnSe crystal scintillator (pure and doped by various elements; 
$\rho=5.65$ g cm$^{-3}$) is very interesting material in which an extremely 
low quenching is observed for $\alpha$ particles: measured values of $Q_\alpha$ are
close to 1. 
For example, $Q_\alpha=1.0\pm0.1$ for 5.5 MeV $\alpha$ particles 
from $^{241}$Am was obtained in \cite{Dan89}.

Several samples of ZnSe doped by O, Al, Cd, Te were investigated in 
Ref. \cite{Lee06}. $Q_\alpha$ values for $\alpha$ particles of $\simeq5.2$ MeV
($^{239}$Pu) were different for crystals with different dopants:
for time of a signal collection $\Delta t=12.8$ $\mu$s, value of
$Q_\alpha=0.70$ was obtained for ZnSe(Cd), and
$Q_\alpha=0.82$ was obtained for ZnSe(O).
For ZnSe(Te), which is a slow scintillator (with decay time for different 
components of scintillating signal as $30-80$ $\mu$s \cite{Lee06}), 
even values of $Q_\alpha>1$ were obtained ($Q_\alpha=1.25$ with $\Delta t=0.6$ $\mu$s,
and 1.13 with $\Delta t=12.8$ $\mu$s). 
Evidently even collection time of $\Delta t=12.8$ $\mu$s is not long enough to
collect total signal in this slow scintillator. We could suppose that 
this is a reason of values $Q_\alpha>1$, and collection of a signal during
longer times would give an usual situation with $Q_\alpha<1$\footnote{Similar
situation with protons which have faster signals than $\gamma$ quanta
in CsI(Tl) was already mentioned in section 2:
$Q_p>1$ was obtained for $\Delta t=1$ $\mu$s, and 
$Q_p<1$ for $\Delta t=7$ $\mu$s.}.

At bolometric temperatures, when signals in ZnSe are extremely long
($>150$ ms), it was possible even to obtain values of $Q_\alpha \simeq4$
\cite{Pir09}.

We have to note here that Eq. (5) supposes {\em quenching} of a signal from
highly ionizing particle in comparison with that from electrons,
and it is impossible to describe any {\em enhancement} on its basis (staying
with physical values of $kB>0$). For $kB=0$, Eq. (5) gives quenching 
factor $Q_i=1$ for any particle. In Fig. 19f, we suppose usual situation when
a signal is totally collected and $Q_i<1$. Quenching factors are drawn
for Zn and Se ions in ZnSe with $kB$ value as:
$kB=1.9\times10^{-5}$ g MeV$^{-1}$ cm$^{-2}$ (which gives   
$Q_\alpha=0.99$ for 5.5 MeV $\alpha$ particles) and
$kB=8.4\times10^{-4}$ g MeV$^{-1}$ cm$^{-2}$ 
($Q_\alpha=0.70$ for 5.5 MeV $\alpha$ particles).

\section{Conclusions}

Semi-empirical and quite simple in realization 
method of calculation of quenching factors for
scintillators was described in this work. It is based on the
classical Birks formula with the {\em total} stopping powers for
electrons and ions, and has only one parameter: the Birks factor
$kB$. Value of this factor for a given
scintillating material can be different in different
conditions of measurements and data treatment. However, if
experimental conditions and treatment of data are fixed, 
hypothesis that $kB$ has the same value for particles of different 
kinds gives reliable results. Once the $kB$ is found by
fitting quenching factors for particles of one kind and in some
range of energies (e.g. for $\alpha$ particles from internal
contamination of a detector by U/Th chains and/or by $^{147}$Sm, 
$^{190}$Pt with energies of a few MeV), 
it can be used to calculate quenching factors for particles
of another kinds and for another energies of interest (e.g. for low energy
nuclear recoils). Many examples were given for materials which,
furthermore, have different mechanisms of scintillation: organic
scintillators (solid C$_8$H$_8$, and liquid C$_{16}$H$_{18}$,
C$_9$H$_{12}$); crystal scintillators (pure CdWO$_4$, PbWO$_4$,
ZnWO$_4$, CaWO$_4$, CeF$_3$, and doped CaF$_2$(Eu), CsI(Tl), 
CsI(Na), NaI(Tl)); and liquid noble gases (LXe). 
It was demonstrated for many cases that
the method allows not only to {\em describe} measured data for
ions of one kind in a reliable way but also to {\em predict}
behaviour of quenching factors for other particles which sometimes
is immediately confirmed by already existing experimental data --
sometimes worse, sometimes better, and sometimes very good, but at least in
a rough agreement.
Some predictions (e.g. for LNe, LiF and others) could be checked in near future.

Stopping powers for electrons and ions are calculated with the
ESTAR and SRIM codes, respectively, which in fact present to-date
state-of-art software in this field. It is easy to use these
programs and they are publicly available; this makes $Q_i$ calculations
quite simple. 

Calculations with the SRIM package have some tendency to overestimate quenching
factors for $\alpha$ particles at energies around $\simeq2$ MeV and underestimate 
them at high energies ($>8$ MeV) as can be seen in
Fig. 4a for CdWO$_4$,
Fig. 5a for CaF$_2$(Eu),
Fig. 8 for CaWO$_4$, and
Fig. 15 for CeF$_3$.
At the same time, calculation of the stopping powers for $\alpha$ particles
with the ASTAR package gave better description of $Q_\alpha$ in CaF$_2$(Eu) 
scintillator at lower energies (see Fig. 5a). 
For some other materials difference between ASTAR and SRIM calculations 
was not big (see Fig. 2 for CsI(Tl) and Fig. 3a for C$_8$H$_8$).
Evidently $Q_i$ values will depend on how one calculates stopping powers
for ions and electrons, and it is a pity that stopping powers could not 
be computed in framework of the same package for any particle 
(SRIM calculates SP for ions in any substance but does not calculate 
SP for electrons; and STAR gives SP for electrons in any 
material but SP for ions are possible only for protons and $\alpha$ 
particles and for a limited list of materials).

Quenching factors calculated in
the presented approach in general increase at low energies, and
this encourages experimental searches for dark matter particles.
Estimations of quenching factors for nuclear recoils are given for some
scintillators where experimental data are absent (CdWO$_4$, PbWO$_4$, CeF$_3$, 
Bi$_4$Ge$_3$O$_{12}$, LiF, ZnSe).

\section{Acknowledgments}

Author is grateful to R. Bernabei, F.A. Danevich and S.K. Kim
for valuable discussions, and to anonymous referee for useful suggestions. 
Work was supported in part by the Brain Pool program of
the Korean Federation of Science and Technology Societies, and 
by the Project ``Kosmomikrofizyka'' (Astroparticle Physics) of the National 
Academy of Sciences of Ukraine.

\end{document}